\documentclass[twocolumn]{IEEEtran}

\usepackage{amsmath,epsfig,amssymb,verbatim,amsopn,cite,subfigure,multirow}
\usepackage{balance}
\usepackage{footnote}
\usepackage{algorithm,algorithmic}
\usepackage[usenames,dvipsnames]{color}
\usepackage[all]{xy}  
\usepackage{url}
\usepackage{adjustbox}
\usepackage{enumitem}
\usepackage{cite,algorithm,algorithmic,amsmath,amssymb,amsthm,empheq,mhsetup}
\usepackage{subfigure,amsfonts,balance}
\usepackage{epstopdf}
\usepackage{setspace}
\usepackage[normalem]{ulem}
\usepackage{acronym}
\usepackage[most]{tcolorbox}
\usepackage[abs]{overpic}
\usepackage{rotating}

\newcommand\hl{\bgroup\markoverwith
    {\textcolor{yellow}{\rule[-.5ex]{.1pt}{2.5ex}}}\ULon}

\newtheorem{Remark}{Remark}

  {\proof}{\proofend}

\DeclareMathOperator{\E}{\mathbb{E}}

\newcommand{\EX}[1]{\E\left\{{#1}\right\}}

\newcommand{\CG}[2]{\mathcal{CN}\left({#1},{#2}\right)}

\newcommand{\cov}[1]{\mathrm{cov}\left({#1}\right)}

\newcommand{\B}[1]{{\mathbf{#1}}}
\newcommand{\C}{\mathbb{C}}

\newcommand{\Pu}{\rho_{\mathrm{u}}}
\newcommand{\Pp}{\rho_{\mathrm{p}}}

\newcommand{\Pd}{\rho_{\mathrm{d}}}

\newcommand{\CH}{\mathrm{CH}}
\newcommand{\FSI}{\mathrm{FSI}}
\newcommand{\ul}{\mathrm{u}}
\newcommand{\p}{\mathrm{p}}
\newcommand{\dl}{\mathrm{d}}

\newcommand{\tauc}{\tau_\mathrm{c}}
\newcommand{\taup}{\tau_\mathrm{p}}
\newcommand{\taud}{\tau_\mathrm{d}}
\newcommand{\tauu}{\tau_\mathrm{u}}

\newfont{\bb}{msbm10 scaled 1100}
\newcommand{\hv}{{\bf g}}
\newcommand{\phiv}{\pmb{\varphi}}
\newcommand{\Sigmam}{\hbox{\boldmath$\Sigma$}}
\newcommand{\CC}{\mbox{\bb C}}

\def\sir {\mbox{\scriptsize\sf SIR}}

\def\sinr {\mbox{\scriptsize\sf SINR}}

\newcounter{mytempeqcounter}

\definecolor{BgYellow}{HTML}{FFF59C}
\definecolor{FrameYellow}{HTML}{F7A600}

\newtcolorbox{StickyNote}[1][]{%
    enhanced,
    before skip=2mm,after skip=2mm, 
    width=\columnwidth, boxrule=0.2mm, 
    colback=BgYellow, colframe=FrameYellow, 
    attach boxed title to top left={xshift=0cm,yshift*=0mm-\tcboxedtitleheight},
    varwidth boxed title*=-3cm,
    boxed title style={frame code={%
        \path[left color=FrameYellow,right color=FrameYellow,
        middle color=FrameYellow]
        ([xshift=-0mm]frame.north west) -- ([xshift=0mm]frame.north east)
        [rounded corners=0mm]-- ([xshift=0mm,yshift=0mm]frame.north east)
        -- (frame.south east) -- (frame.south west)
        -- ([xshift=0mm,yshift=0mm]frame.north west)
        [sharp corners]-- cycle;
        },interior engine=empty,
    },
    sharp corners,rounded corners=southeast,arc is angular,arc=3mm,
    underlay={%
        \path[fill=BgYellow!80!black] ([yshift=3mm]interior.south east)--++(-0.4,-0.1)--++(0.1,-0.2);
        \path[draw=FrameYellow,shorten <=-0.05mm,shorten >=-0.05mm,color=FrameYellow] ([yshift=3mm]interior.south east)--++(-0.4,-0.1)--++(0.1,-0.2);
        },
    drop fuzzy shadow, 
    fonttitle=\bfseries, 
    title={#1}
}

\begin{document}
\title{Ultra-Dense Cell-Free Massive MIMO for 6G: Technical Overview and Open Questions}

\author{Hien Quoc Ngo,~\IEEEmembership{Senior Member,~IEEE,} Giovanni Interdonato,~\IEEEmembership{Member,~IEEE,}
Erik G. Larsson,~\IEEEmembership{Fellow IEEE},  Giuseppe Caire,~\IEEEmembership{Fellow IEEE}, and Jeffrey G. Andrews,~\IEEEmembership{Fellow IEEE}\\ (\textit{Invited Paper})
\thanks{H. Q. Ngo is with the Centre for Wireless Innovation (CWI), Queen's University Belfast, U.K. (hien.ngo@qub.ac.uk).
}
\thanks{G. Interdonato is with the Department of Electrical and Information Engineering, University of Cassino and Southern Lazio, Cassino, Italy (giovanni.interdonato@unicas.it).}

\thanks{E. G. Larsson is with the Department of Electrical Engineering (ISY), Link\"{o}ping University, Sweden (e-mail: erik.g.larsson@liu.se).}

\thanks{Giuseppe Caire is with the Faculty of Electrical Engineering and Computer Science, Technical University of Berlin,  Germany (e-mail: caire@tu-berlin.de).}

\thanks{Jeffrey G. Andrews is with 6G@UT in the Wireless Networking and Communications Group, Dept. of ECE, The University of Texas at Austin, USA.  (email: jandrews@ece.utexas.edu).
}
\thanks{The work of H. Q. Ngo was supported by the U.K. Research and Innovation Future Leaders Fellowships under Grant MR/X010635/1, and a research grant from the Department for the Economy Northern Ireland under the US-Ireland R\&D Partnership Programme.  The work of E. G. Larsson was supported by the Knut and Alice Wallenberg Foundation, the Swedish Research Council, and ELLIIT.}
}

\allowdisplaybreaks

\bstctlcite{IEEEexample:BSTcontrol}
\maketitle

\begin{abstract}
Ultra-dense cell-free massive multiple-input multiple-output (CF-MMIMO) has emerged as a promising technology expected to meet the future ubiquitous connectivity requirements and ever-growing data traffic demands in 6G. This article provides a contemporary overview of ultra-dense CF-MMIMO networks, and addresses important unresolved questions on their future deployment. We first present a comprehensive survey of state-of-the-art research on CF-MMIMO and ultra-dense networks. Then, we  discuss the key challenges of CF-MMIMO under ultra-dense scenarios such as low-complexity architecture and processing, low-complexity/scalable resource allocation, fronthaul limitation, massive access, synchronization, and channel acquisition. Finally, we answer key open questions, considering different design comparisons and discussing suitable methods dealing with the key challenges of ultra-dense CF-MMIMO. The discussion aims to provide a valuable roadmap for interesting future research directions in this area, facilitating the development of CF-MMIMO for 6G.
\end{abstract}

\begin{IEEEkeywords}
Cell-free massive MIMO, network MIMO, open RAN, ubiquitous connectivity, ultra-dense networks, user-centric sytems,  6G.  
 \end{IEEEkeywords}

\section{Introduction} 

\begin{figure}[t!]
\centerline{\includegraphics[width=0.45\textwidth]{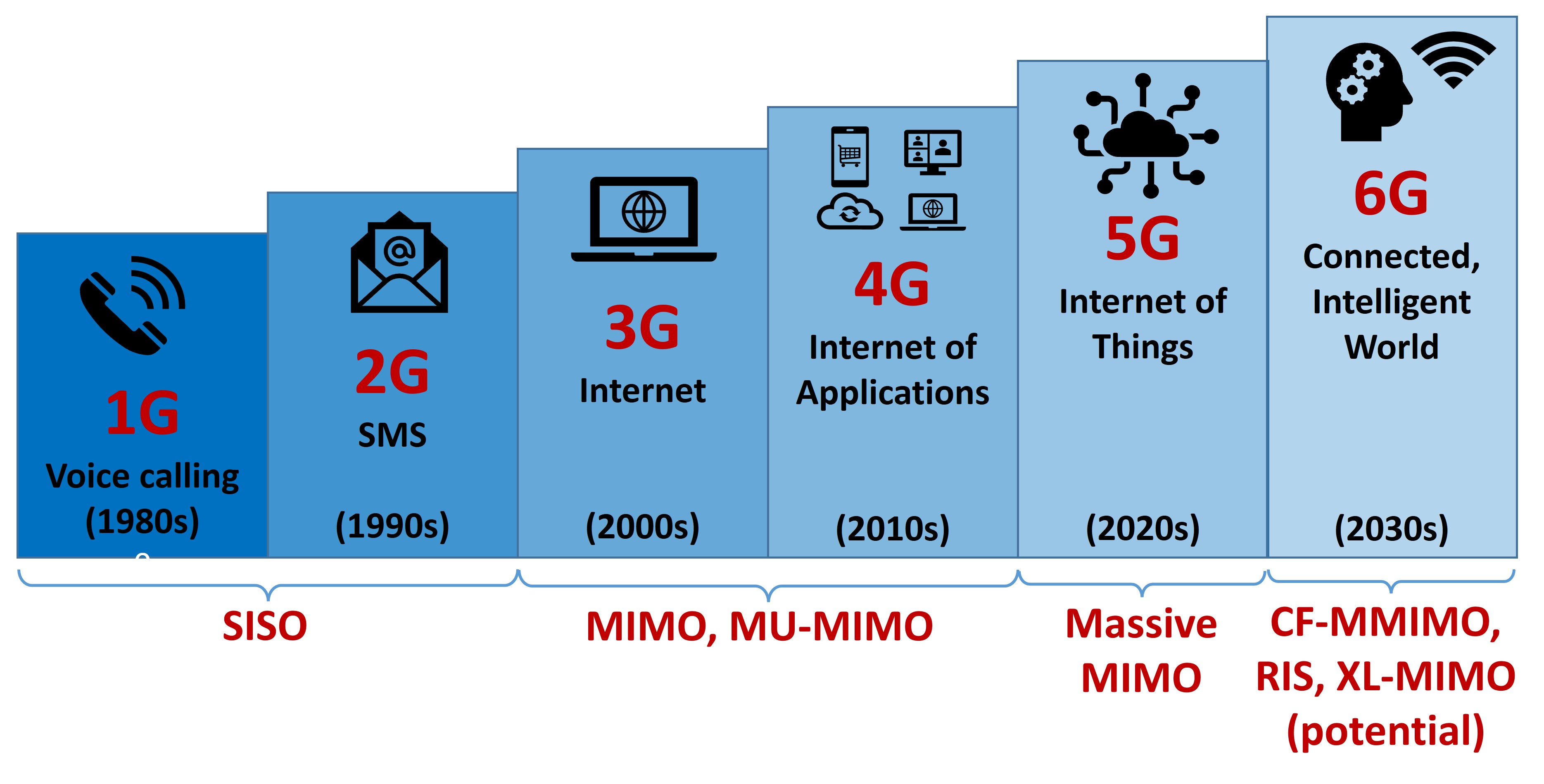}}
\caption{Evolution and paradigm shifts of mobile networks.\label{fig:mobilegeneration}}
\end{figure}

Mobile wireless communication systems have undergone rapid evolution, with operators and regulators rolling out a new generation of mobile technology approximately every decade.  Each generation, spanning from the first to the fifth, has been meticulously designed to address the evolving requirements of both end-users and network operators \cite{ohmori2000future,adachi2001wireless,liu2017nonorthogonal}; see Figure~\ref{fig:mobilegeneration}. The first generation (1G) was based on analog communications, and deployed in the 1980s. These systems were basically used for voice calls only. They had poor performance with poor voice quality, only one call per channel, and low capacity. In 1G, frequency division multiple access (FDMA) was adopted. The second generation (2G) of wireless systems was introduced in the 1990s. Different from 1G, 2G was based on digital technology in the cellular environment with different standards such as global system for mobile (GSM), code division multiple access one (CDMA-One), IS-136, and pacific digital cellular (PDC). The 2G systems offered  much better performance, better reliability and higher capacity, and supported text messages. The third generation (3G) was introduced in 2000 and provided multimedia support. The initial phase of 3G was considered as 2.5G. In 3G, data was added along with voice and general packet radio service (GPRS) was introduced which provided services such as e-mail and picture messages. The 4G was introduced in 2014 to support wireless mobile Internet with current and emergent multimedia services such as video chat, digital video broadcast (DVB), mobile TV, and high-definition television (HDTV). It can provide peak data rates in the range of 100 Mbps to 1 Gbps, and a latency between 40 milliseconds (ms) and 60 ms. The famous standards used are long term evolution-advanced (LTE-A) by the 3rd generation partnership project (3GPP). 

With the aim of fully implementing the concept of ``anywhere'' and ``anytime'', as well as to support new, diverse, and emergent services, users are demanding more and more from the cellular networks. New requirements include increasing throughputs, lower delays, and enhanced spectrum efficiency. These are key requirements necessary to deliver the new emergent broadband data services. In order to meet these requirements, 5G was introduced in 2020 \cite{parkvall20205g,liu20205g}. 5G is expected to support enhanced mobile broadband (eMBB), ultra-reliable low latency communications (URLLC), and massive machine-type communications (mMTC) with specific requirements: (i) high capacity for eMBB (up to 10~Gbps for uplink (UL) and 20~Gbps for downlink (DL) with a bandwidth of 400~MHz)  and 100 times more  connected devices than in 4G; (ii) URLLC with less than $10^{-5}$
block error rate and 1 ms or less latency; and (iii) mMTC implementation and incorporation of low-power wide area network (LPWAN) requirements to strengthen IoT solutions \cite{series2015imt}.

The application trends foreseen in the next decade include a wide range of advancements such as ubiquitous intelligence and computing, smart X, immersive extended reality, ubiquitous connectivity, the seamless integration of sensing and communications, as well as holographic-type communications. These trends signify a growing demand for a more advanced network infrastructure, fulfilling the future needs of both individual users and various vertical industries. Consequently, though 5G deployment is still progressing across the world, industry and academia have already proactively initiated the conceptualization of beyond 5G, commonly referred to as the sixth generation (6G) systems \cite{jiang2021road,matthaiou2021road,uusitalo20216g,wang2023road}. 6G  is expected to be operating  around 2030.
In contrast to 5G, 6G is anticipated to not only continue supporting existing 5G usage scenarios (i.e., eMBB, URLLC, and mMTC) but also to support enhanced and novel capabilities \cite{series2023imt,liu2023beginning}. Furthermore, 6G is poised to enable new use cases stemming from advanced capabilities, particularly those related to artificial intelligence (AI) and multi-sensory interactions. Notably, ubiquitous connectivity is a central focus of 6G where the connection density is expected to be $10^8$ devices per km$^2$, exceeding 5G's primary focus by a factor of $100$ \cite{series2023imt}. 

Ultra-dense cell-free massive multiple-input multiple-output (CF-MMIMO) stands out as a promising wireless networking technology capable of addressing the ubiquitous connectivity requirements and ever-growing demands for data traffic in 6G. In ultra-dense CF-MMIMO, many access points (APs), distributed over a wide area, coherently serve many users. The total number of all APs' antennas is much larger than the number of users that are  scheduled jointly  in the same time-frequency resource. In other words, ultra-dense CF-MMIMO is an ultra-dense network (UDN) with AP cooperation and massive MIMO-based techniques. This system can be considered as the combination of UDNs, massive MIMO, and network MIMO (a.k.a. coordinated multipoint joint transmission (CoMP-JT) systems). Thus it can reap all benefits from these systems and provide ubiquitous connectivity with very high spectral and energy efficiency \cite{ngo2015cell,Hien:cellfree,Hien:TGCN:2018,interdonato2019ubiquitous,elhoushy2021cell,demir2021foundations}.

While the theoretical foundations of CF-MMIMO are well-established, the practical implementation of this technology remains very challenging, particularly in UDNs undergoing increased densification \cite{demir2021foundations,Zhang:Access:2019,Shuaifei:DCN:2022}. There are still many open questions that must be addressed before rolling out this technology in practice. This paper aims to bridge this gap by first presenting a comprehensive survey of ultra-dense CF-MMIMO. Then, we discuss key questions related to practical implementation of CF-MMIMO, considering computational complexity and performance tradeoffs. Our exploration includes various aspects such as practical infrastructure, signal processing, capacity bounding techniques, massive access, resource allocation, synchronization and calibration. Note that, while CF-MMIMO can be implemented across various frequency bands, our focus herein is on sub-6G bands. This focus comes from the anticipation that these frequency bands will continue serving as primary carriers for handling the traffic in 6G. Moreover, the propagation characteristics of sub-6G frequencies are significantly superior to those of mmWave or higher bands, thereby unlocking the full potential of CF-MMIMO.

\begin{table}[t!]
    \caption{
        List of Abbreviations.
    }
    \centerline{\small
\begin{tabular}{|l|l|}
  \hline
  \textbf{Abbreviations} & \textbf{Definition} \\ \hline
  3GPP & Third generation partnership project\\ \hline
  5G & Fifth generation \\ \hline
  6G & Sixth generation \\ \hline
  AI & Artificial intelligence \\ \hline
  AoA & Angle of arrival \\ \hline
  AoD & Angle of departure \\ \hline
  AP & Access point \\ \hline
  APU & Antenna processing unit \\ \hline
  BS & Base Station \\ \hline
  C-MMSE & Centralized MMSE \\ \hline
  CF-MMIMO & Cell-free massive MIMO \\ \hline
  CoMP-JT & Coordinated multipoint joint transmission \\ \hline
  CPU & Central processing unit \\ \hline
  CSI & Channel state information \\ \hline
  DCC & Dynamic cooperation clustering \\ \hline
  DL & Downlink \\ \hline
  DNN & Deep neural network \\ \hline
  DRL & Deep reinforcement learning \\ \hline
  eMBB & Enhanced mobile broadband \\ \hline
  FDD & Frequency-division duplex \\ \hline
  L-MMSE & Local MMSE \\ \hline
  LP-MMSE & Local partial MMSE \\ \hline
  LoS & Line of sight \\ \hline
  MIMO & Multiple-input multiple-output \\ \hline
  MMF & Max-min fairness \\ \hline
  MMSE & Minimum mean-square error \\ \hline
  mMTC & Massive MTC \\ \hline
  MR & Maximum-ratio \\ \hline
  MTC & Machine-type communication \\ \hline
  MU-MIMO & Multi-user MIMO \\ \hline
  NR & New radio \\ \hline
  O-DU & O-RAN distributed unit \\ \hline
  O-RAN & Open RAN \\ \hline
  O-RU & O-RAN radio unit \\ \hline
  OTA & Over-the-air \\ \hline
  PHY & Physical layer \\ \hline
  PZF & Partial ZF \\ \hline
  QoS & Quality of service \\ \hline
  RAN & Radio access network \\ \hline
  SE & Spectral efficiency \\ \hline
  SNR & Signal-to-noise ratio \\ \hline
  SINR & Signal-to-interference-plus-noise ratio \\ \hline
  TMMSE & Team MMSE \\ \hline
  TDD & Time-division duplex \\ \hline
  UDN & Ultra-dense network \\ \hline
  UL & Uplink \\ \hline
  URLLC & Ultra-reliable low-latency communications\\ \hline
  ZF & Zero-forcing \\ \hline
\end{tabular}}
    \label{table:1}
\end{table}

The rest of this paper is organized as follows. In Sections~\ref{Sec:CFmMIMO_SoA} and \ref{Sec:UDN_SoA}, we provide a comprehensive overview of CF-MMIMO and UDNs as well as their challenges. In Section~\ref{sec:systemmodel}, we present a general mathematical model for user-centric CF-MMIMO. We discuss open questions in Section~\ref{sec:questions} and finally conclude the paper
in Section~\ref{sec:conclusions}.

\textit{Notation}: We use bold upper-case and lower-case letters to denote matrices and vectors, respectively. The superscripts $(\cdot)^T$, $(\cdot)^*$, and $(\cdot)^H$ stand for the transpose, conjugate, and conjugate-transpose operations, respectively.   $\B{I}_M$  represents the $M\times M$ identity matrix, while $|\cdot|$ is the cardinality of a set. The  determinant and the Euclidean norm operations are denoted by  $\text{det}(\cdot)$, and $\| \cdot\|^2$. 
The notation $\mathcal{CN}(0,\sigma^2)$ represents circular symmetric complex Gaussian distribution having variance $\sigma^2$. Finally, $\EX{\cdot}$ and $\cov{\cdot}$ denote the statistical expectation and covariance matrix. 

\textit{Abbreviations}: A list of abbreviations is presented in Table~\ref{table:1}.

\section{Cell-Free Massive MIMO: State of the Art } \label{Sec:CFmMIMO_SoA}

The advent of successive wireless generations has spurred the evolution of multiple-antenna technologies, see Figure~\ref{fig:mobilegeneration}. Whereas first and second generation wireless networks utilized single-input single-output (SISO) systems, third and fourth generation networks adopted multiple-input multiple-output (MIMO) and multi-user MIMO (MU-MIMO) technologies. In the current 5G landscape, massive MIMO, a form of MU-MIMO with a large number of antennas at the base station (BS), is considered a core technology. Furthermore, CF-MMIMO, which operates without cellular topology, is expected to be a fundamental component of 6G wireless networks. In the subsequent discussion, we will provide details of multiple antenna technology, ranging from MIMO to CF-MMIMO.

\subsection{Point-to-Point MIMO}\label{sec:MIMO}

MIMO technology uses multiple antennas for both transmitter and receiver. The MIMO concept was developed in the 1990s \cite{paulraj1994increasing,raleigh1998spatio,Fosc98}, and since then, it has found widespread application in various wireless local area networks (WLANs) and evolving cellular communications standards. It stands as one of the key technologies in LTE Rel-8. There are two major operations under MIMO: transmit diversity and spatial multiplexing.  With transmit diversity, multiple antennas simultaneously transmit the same data to enhance the signal-to-noise ratio (SNR). In contrast, with spatial multiplexing, the transmit antennas send independent data streams  to increase capacity. 

\begin{figure*}[t]
	\centering
  \vspace{0em}
	\includegraphics[width=160mm]{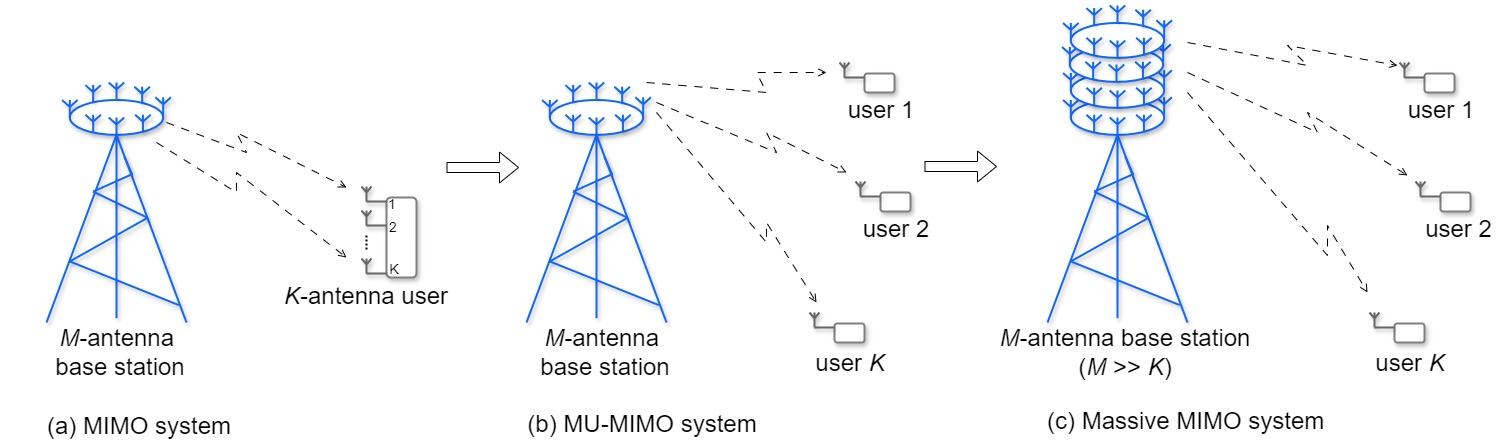}
		\caption{Multiple antenna technology: from MIMO to massive MIMO.} 
 \vspace{-0.5em}
	\label{fig:MutipleAntennas}
\end{figure*}


A point-to-point system is shown in Figure~\ref{fig:MutipleAntennas}(a) where a BS having $M$ antennas transmits data to a user of $K$ antennas. The propagation channel between the BS antennas and the user antennas is an $M\times K$ matrix of complex-valued coefficients, denoted by $\B{G}$. We assume block-fading channels where the channel is static during each coherence interval/block (i.e. coherence time times coherence bandwidth), and independently varies between different coherence intervals. With perfect channel state information (CSI) knowledge at the user, an (instantaneous) achievable rate (in bits/s/Hz) of the DL transmission is given by
\begin{align}\label{eq:DLMIMO_capacity1}
    R &=  \log_2\det\left(\B{I}_K + \frac{\Pu}{M} \B{G}^H \B{G} \right),
\end{align}
where $\Pu$ is the receive SNR, i.e. the power divided by the noise-variance. The above rate is achieved under uncorrelated coded data streams with equal power on all the transmit antennas.

If the channel $\B{G}$ has i.i.d. $\CG{0}{1}$ elements, then at high SNR, the achievable rate \eqref{eq:DLMIMO_capacity1} can be approximated as
\begin{align}\label{eq:DLMIMO_capacity2}
    R &\approx  \min(M,K)\log_2 \Pu.
\end{align}
We can see that, at high SNR, the achievable rate scales as $\min(M,K)$ -- the multiplexing gain. We can increase the system rate by just adding more antennas at the BS and the user.\footnote{If the BS has knowledge of the CSI, with specific linear precoding and power allocation over the channel eigenmodes, we can achieve a higher rate, compared to the rate in \eqref{eq:DLMIMO_capacity1}. However, the high-SNR behavior mentioned above remains unchanged \cite{foschini1996layered}.} 
However this is true only when the SNR is high and the channel is full rank. In low-SNR scenarios (e.g. users at the cell edges) or/and low-rank propagation channels (e.g. line-of-sight (LoS), double scattering channels, keyhole channels), we cannot leverage the advantage of the multiplexing gain in MIMO. This point is illustrated by the following example. Consider a geometric channel model with $L$ paths between the BS and the user, and assume that both the BS and the user are equipped with uniform linear arrays  with half-wave length antenna spacing. Then, the channel $\B{G}$ can be modeled as  \begin{align}\label{eq:channel_0}
    \B{G} = \frac{1}{\sqrt{L}}\sum_{l=1}^L \alpha_l \B{a}\left(\theta_l\right)\B{b}\left(\phi_l\right)^H,
\end{align}
where $\alpha_l$, $\theta_l$, and $\phi_l$ are the complex path gain, angle-of-departure (AoD) and angle-of-arrival (AoA), respectively, of path $l$. In addition, $\B{a}\left(\theta_l\right)\in \mathbb{C}^{M\times 1}$ and $\B{b}\left(\phi_l\right)\in \mathbb{C}^{K\times 1}$ are the array response vectors at the BS and the user, respectively, given by
\begin{align}\label{eq:channel_1}
    \B{a}\left(\theta_l\right) &=\left[1 ~ e^{j \pi \sin(\theta_l)} \ldots e^{j(M-1) \pi \sin(\theta_l)} \right]^T,\\
    \B{b}\left(\phi_l\right) &=\left[1 ~ e^{j \pi \sin(\phi_l)} \ldots e^{j(K-1) \pi \sin(\phi_l)} \right]^T.
\end{align}

Clearly, $\text{rank}\left(\B{G}\right) \leq \min(M,K,L)$, which is limited by $L$ even when $M$ and $K$ are large. The rank of the channel matrix represents the number of spatial degrees of freedom of the channel, which in turn determines the number of independent data streams we can
send, with rate that scales as $\log_2(\text{SNR})$. These results are asymptotically tight at high
SNR in the sense that they match information theoretic bounds. In general, the higher the channel rank, the more independent data streams can be sent, and hence, we can obtain higher rate. Note that if the angles are not well separated, some non-zero eigenvalues are small, and the actual number of dominant data streams is even smaller than channel matrix rank. In an extreme case where $L=1$, we have a LoS channel with a rank-one matrix, and the rate \eqref{eq:DLMIMO_capacity1} becomes
\begin{align}\label{eq:DLMIMO_capacity_LoS}
    R &=  \log_2\left(1 + \Pu K \left|\alpha_1\right|^2 \right).
\end{align}
 In this case, one spatial degree of freedom is available. Increasing  $K$ can improve the SNR at the user,  but it does not improve the degrees of freedom/multiplexing gain.

\subsection{Multiuser MIMO}\label{sec:MUMIMO}

MU-MIMO was introduced in the 2000's \cite{viswanath2003sum,caire2003achievable},   building upon the original SDMA/TDMA concept for satellite communications introduced in the 1970s \cite{tsuji1976transmit},  as an innovative solution to address some of the drawbacks of point-to-point MIMO, discussed in Section~\ref{sec:MIMO}. Figure~\ref{fig:MutipleAntennas}(b) shows a MU-MIMO system model, in which a multiple-antenna BS serves multiple users, each equipped with either one or several antennas. By enabling multiple users to be co-scheduled on the same time-frequency resources, MU-MIMO can achieve the same spatial multiplexing gain as the point-to-point MIMO, even when users have only a single antenna. This is particularly advantageous as the users are constrained by the number of antennas and by antenna design considerations. Furthermore, MU-MIMO demonstrates resilience to propagation due to the distributed nature of the users. It still works well in scenarios with low-rank channels. For example, consider again the case of LoS channels. In LoS, the channel vector between the BS and the $k$-th user can be represented as 
\begin{align}\label{eq:MU_MIMO_channel_1}
    \B{g}_k=\alpha_k \B{a}(\theta_k),
\end{align}
where  $\theta_k$ is the AoA for the $k$-th user. Clearly, if $\theta_k\neq \theta_{k'}$, $\forall k\neq k'$ (this is nearly certain to happen, as  users are distributed randomly within the coverage area), the channel matrix $\B{G} = [\B{g}_1 \ldots \B{g}_K]$ is full rank, i.e. $\text{rank}(\B{G})=\min(M, K)$. Therefore, a full spatial degrees of freedom is achieved. However, the end-to-end performance  actually depends on the number of dominant singular values. Thus, to achieve high rates, the AoAs to different users have to be well separated.


Though the information theoretic capacity region can be achieved by using non-linear precoding (i.e. dirty paper coding) in the DL, and non-linear successive interference cancellation in the UL, simple linear processing such as zero-forcing (ZF) or minimum mean-square
error (MMSE) processing achieves excellent performance \cite{yoo2006optimality,boccardi2007near}. Thus MU-MIMO is well adopted in practice. However, early proposals for MU-MIMO implementations considered cellular systems operating in frequency-division duplex (FDD), as in most deployed 4G systems, or evolutions of WiFi (specifically, 802.11ax), for which, due to various reasons, channel reciprocity does not hold. Thus, in order to compute the precoder at the BS, typically the DL channel is probed via a DL pilot signal and CSI is fed back from the users to the BS. This requires a DL pilot signal dimension at least equal to the number of BS antennas. Since one round of DL pilot transmission and UL feedback is performed for each channel coherence time-frequency block, the number of BS antennas cannot be made arbitrarily large; otherwise, the DL pilot dimension would consume the whole coherence block. Due to this limitation, the initial proposals for MU-MIMO focused on a relatively small number of BS antennas. 
This faces challenges in simultaneously serving  many users  in the same frequency band, and hence, limits the suitability of conventional MU-MIMO in  dense networks.

\subsection{Massive MIMO}

Massive MIMO technology emerged in the early 2010s, and was introduced by Thomas L. Marzetta in his seminal works \cite{Mar:06:ACSSC,Mar:10:WCOM}. It presents a scaled-up version of MU-MIMO. It is worth noting that while \cite{Mar:06:ACSSC,Mar:10:WCOM} did not explicitly employ the term ``massive MIMO'', it provided the fundamental underpinnings that catalyzed the subsequent exploration of this concept. Over the years, massive MIMO has had various names, including large-scale antenna systems, very large MU-MIMO, and full-dimensional MIMO \cite{Marzetta:JSAC:2013,NLM:13:TCOM,li2013implementation}. Nowadays, the massive MIMO term has been widely adopted both within academia and industry.
%

In massive MIMO, the BSs are equipped with many antennas, allowing them to spatially multiplex many users  on the same time-frequency resource; see Figure~\ref{fig:MutipleAntennas}(c). Typically, the number of BS antennas exceeds that of all users' antennas by several times. Key aspects of massive MIMO are: 1) an excess of service antennas relative to the number of streams; and 2) the operation in time division duplex (TDD), relying on uplink-downlink channel reciprocity. The reciprocity is the most important enabling factor of massive MIMO: the UL channels are estimated at the BS via UL pilots, and the DL channels are determined based on the channel reciprocity. As a result, the channel estimation overhead depends only on the number of users,  irrespective of  the number of BS antennas. Any number of BS antennas can be trained by a finite number of pilots. This makes the system scalable in the sense that adding more antennas to the BS does not cost UL pilot resources, and always enhance the system performance. We can say that massive MIMO is MU-MIMO ``done right''.

With numerous antennas at the BS, massive MIMO can offer high array and multiplexing gains. Consequently, it can provide huge energy efficiency and spectral efficiency (SE). Moreover,  by leveraging favorable propagation and channel hardening characteristics, simple processing (e.g., linear precoding and combining, channel statistics-based resource allocation) can yield excellent performance, making  massive MIMO highly practical for implementation \cite{ngo16}. Note that favorable propagation refers to scenarios where the channel vectors from the  BS to different users are pairwisely orthogonal, while channel hardening implies that the squared norms of the channel vectors become deterministic \cite{ngo2017no,emil17}.

Massive MIMO is a remarkable innovation, and has  evolved into the core physical-layer   technology in 5G. The first phase of 5G New Radio (NR) with massive MIMO was standardized by the 3GPP. Nevertheless, existing massive MIMO systems are based on a cellular topology, wherein the geographical coverage area is partitioned into cells, and each cell is served by  one BS equipped with  many antennas. With this cellular-based structure, massive MIMO encounters challenges in ensuring robust connectivity throughout the entire network. This issue arises from the boundary effect, an inherent limitation of cellular-based networks. 

As we show in the next two sections, CF-MMIMO addresses the above limitation of cellular-based massive MIMO systems by building upon the paradigm of Network MIMO.

\subsection{Network MIMO and Base Station Cooperation}

\begin{figure}[t!]
\centering
\subfigure[Conventional cellular system.]{%
\includegraphics[width=0.40\textwidth]{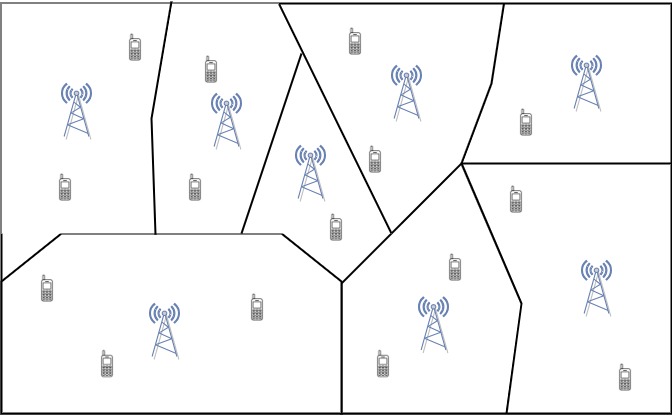}
\label{fig:cellular1}} \quad
\subfigure[Network-centric system.]{%
\includegraphics[width=0.40\textwidth]{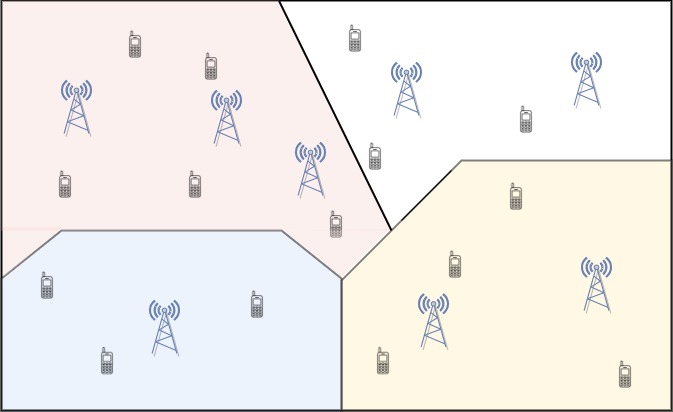}
\label{fig:network-centric1}} 
\quad
\subfigure[User-centric system.]{%
\includegraphics[width=0.40\textwidth]{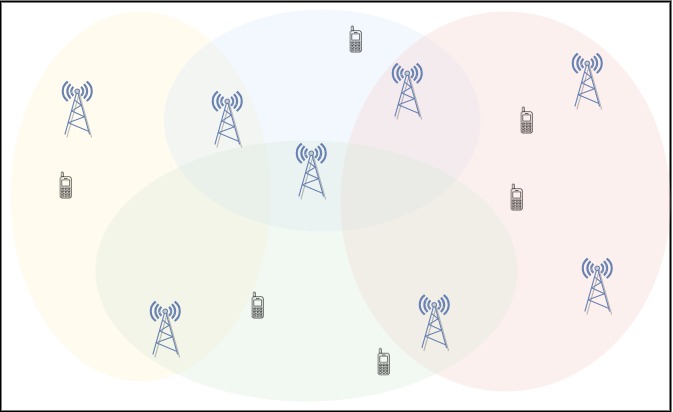}
\label{fig:user-centric1}} 
\caption{Different cellular architectures from lowest (top) to highest (bottom) amounts of cooperation between the BSs.} \label{fig:network-deployment}
\end{figure}

Conventional mobile networks are based on a cellular structure, see Figure~\ref{fig:cellular1}.  An inherent limitation of cellular networks is inter-cell interference, which arises from BSs in other cells (DL) or the users in neighboring cells (UL). This issue is particularly pronounced for DL users near the cell boundaries, as they experience poor signal-to-interference-plus-noise ratio (SINR) due to the  effects of strong inter-cell interference and high path loss from their own BS. 

Since its inception in the 1970s \cite{young1979advanced}, the cellular wireless network has undergone a series of transformative innovations aimed at reducing the boundary effect bottleneck. One notable step forward, known as network MIMO,\footnote{Network MIMO has many different subcategories, and has gone by many alternative names that are nearly synonymous, including BS cooperation, coordinated multi-point (CoMP),  distributed MIMO, and cloud radio access network (CRAN).} was first introduced in 1994 \cite{wyner1994shannon}, with a good summary of the early work appearing in \cite{Gesbert2010a}.  Network MIMO refers to the setup where BSs cooperate over the core network to create a distributed MIMO transmission system, in which a number of nearby BSs pool their antennas to create an effectively much larger MIMO array.   

In the original form of network MIMO, the geographical coverage area is segmented into fixed disjoint clusters, each  jointly served by multiple BSs. This arrangement is commonly referred to as a network-centric topology, as illustrated in Figure~\ref{fig:network-centric1}. Network-centric topologies 
have the advantage of fixed connectivity for the cluster-based processing. However, network MIMO can also be conceptualized as a cellular network with expanded cell sizes, with each expanded ``cell'' being served by multiple BSs. Nonetheless, the notions of cells and cell boundaries persist, making inter-cell interference  a  fundamental limitation \cite{Lozano2013a}. 

An alternative approach to network MIMO replies on user-centric cell formation, where each user is served by its specifically chosen APs, known as a user-centric cluster \cite{tolli2008cooperative,garcia2010dynamic}. As such, the cells will be dynamically determined depending on user locations, see Figure~\ref{fig:user-centric1}. Thus, the boundary effect in cellular networks can be mitigated. However, given the dynamic nature of user-centric clusters, user-centric systems require a flexible fronthaul infrastructure capable of dynamically routing fronthaul traffic. 

In general, the implementation of network MIMO technology necessitates complicated signal co-processing procedures, accompanied by substantial fronthaul/backhaul overheads and deployment costs, all aimed at mitigating the effect of inter-cell interference. The 3GPP LTE standard made an initial attempt to establish standardized cooperation among BSs with the introduction of the CoMP-JT technique, which has evolved to a slightly more general notion of Multiple Transmission Points (Multi-TRP) in 5G NR. Nevertheless, CoMP-JT has not performed very well, in large part due to poor CSI quality (especially for the DL) and insufficiently fast communication and  synchronization amongst the networked BSs. As a result, this technique struggled to gain wide deployment or impact \cite{fantini2016coordinated}.

\subsection{Cell-Free Massive Massive MIMO}

The concept of CF-MMIMO was initially introduced in 2015  with the objective of mitigating the inherent boundary effect of cellular-based systems, achieved through the elimination of traditional cellular cells \cite{ngo2015cell,Hien:cellfree}.  

In general, CF-MMIMO refers to a system comprising a large number of APs and active\footnote{With the terminology \textit{active} user, we mean a user with an established connection to at least one AP. In standards (LTE/NR) 
terminology, this status would correspond to the \textit{radio resource control (RRC) connected}.} users distributed over a large area.  Unlike traditional cellular systems, CF-MMIMO does not have fixed cells or cell boundaries. 
 Canonical CF-MMIMO, as first  introduced in \cite{ngo2015cell}, represents a system where all APs collaborate in serving all users by leveraging a single CPU through fronthaul connections. This canonical configuration serves as an idealized model, offering valuable insights and advantages of CF-MMIMO systems. Nevertheless, it lacks scalability, i.e. making it impractical as the network size (number of APs and/or number of users) grows large.  More practical and scalable approach involves each user being served by a select subset of APs, rather than  all APs, in a user-centric manner. In this user-centric approach, each user chooses a subset of APs which will serve it. By doing this, the users create their own cells (a.k.a. the user-centric clusters), and these cells dynamically change depending on user load and locations. There is no cell boundary as in cellular-based networks. Thus, CF-MMIMO can be considered as an upscaled version of user-centric network MIMO (as shown in  Figure~\ref{fig:user-centric1}) with three specific traits:
 \begin{enumerate}
    
 \item The APs use directly measured CSI on both the UL and DL to operate phase-coherently with full digital processing, to serve all users in the same time-frequency band.
 
 \item TDD transmission is employed, with channel reciprocity assumed. More precisely, in each coherence interval   three main activities take place: UL training for channel acquisition at the APs, UL payload transmission and reception wherein the APs used their channel estimates to combine and detect all the received signals, and DL payload transmission whereby the channels estimated during the UL training phase are used to precode and beamform the data to all users in a combined fashion. The TDD operation is the real enabling factor of the CF-MMIMO  as it entails no common DL probing and CSI feedback, thereby preventing many complications, such as expensive estimation overhead, and stale CSI due to feedback delays.
 
\item The system relies on an excess of service antennas (total number of antennas from all APs) relative to the number of served streams. This provides a significantly greater degree of freedom in managing interference, resulting in substantial performance improvements as opposed to conventional network MIMO. One of the reasonable settings for CF-MMIMO is when the number of APs is less than the number of users, but the total number of antennas from all APs is much greater than the number of users.
    
 \end{enumerate}

 \begin{figure}[t!]
\centering
\subfigure[Co-located massive MIMO.]{%
\begin{overpic}[width=0.48\textwidth]{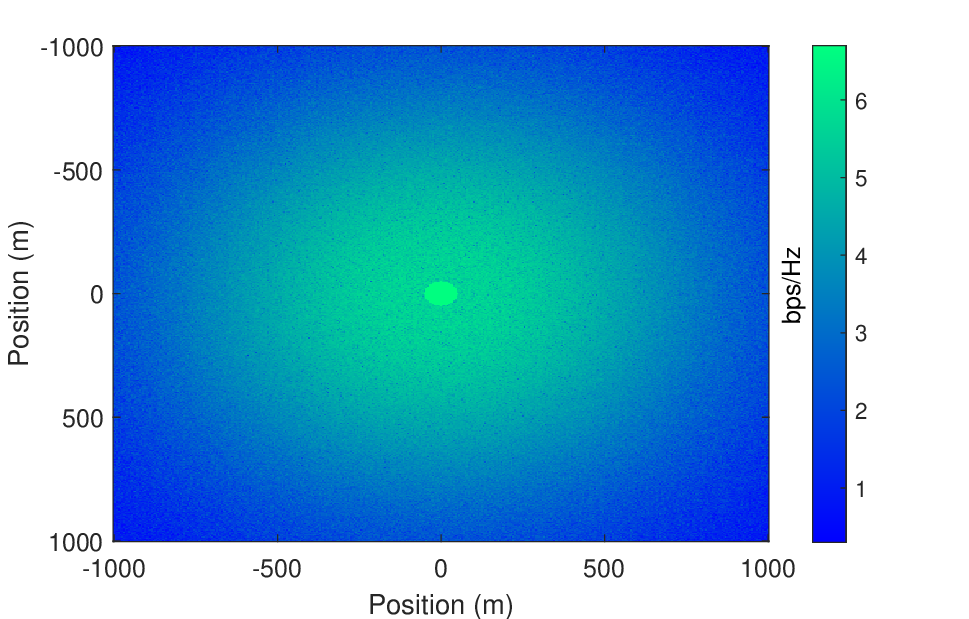}
\end{overpic}
\label{fig:colocated}} \quad
\subfigure[Cell-free massive MIMO.]{%
\begin{overpic}[width=0.48\textwidth]{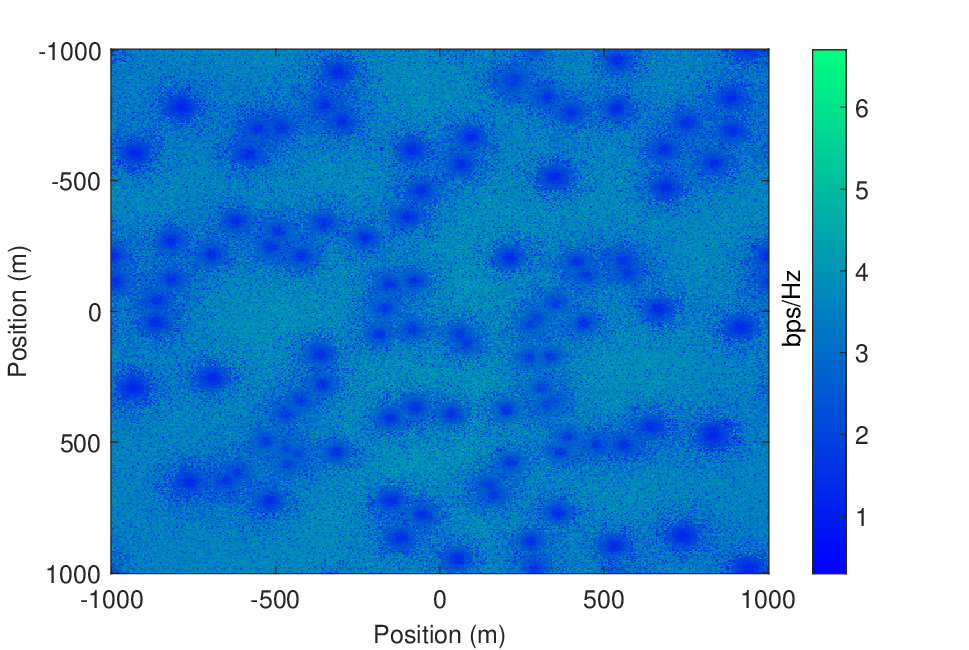}
\end{overpic}
\label{fig:cellfree}} \caption{The achievable rates in bps/Hz for different user locations displayed with scaled colors. Here, for CF-MMIMO, $100$ APs are uniformly located at random in a $2~\text{km}\times 2~\text{km}$ area; while for co-located massive MIMO, all 100 service antennas are located at the center of the square area. } \label{fig:cell-freevscolo}
\end{figure}
 
The above features distinguish CF-MMIMO from other technologies such as small-cell and conventional user-centric network MIMO discussed in the prior section. Note that, when we use the term ``cell-free'', we simply refer to the data plane, i.e., there is no fixed cells created by the data transmission protocol in active mode. Numerous specific mechanisms beyond the Physical Layer (PHY) data plane remain undefined or inadequately defined. Consider, for instance, the processes of user clustering or the initial connection and authentication of new users. These procedures may align with the principles of current cellular 5G-NR. Therefore, even in a ``cell-free'' network, the underlying cellular architecture may persist. In the future, the control plane may also adopt a cell-free approach. However, this presents considerable challenges as it depends on factors that are predominantly standard-dependent rather than purely theoretical/technical.

The authors of \cite{Hien:cellfree} demonstrated that CF-MMIMO could provide consistently high-quality service to all users with very simple maximum-ratio (MR) processing. This can be verified via Fig.~\ref{fig:cell-freevscolo} where the  rates displayed with scaled colors for cell-free and co-located massive MIMO are presented. We can see that, compared to co-located massive MIMO, cell-free systems can offer a much more uniform connectivity for all users in any location. Moreover, it outperforms the conventional small-cell systems. In addition, in LoS propagation environments, CF-MMIMO with multiple-antenna users can offer favorable propagation, but co-located massive MIMO may not. This enables multi-stream transmission even in line-of-sight, a unique advantage of CF-MMIMO as compared to co-located massive MIMO. As a result, CF-MMIMO  has attracted significant attention and interest within the research community ever since its inception.

The literature on CF-MMIMO is rich. Early works on CF-MMIMO focused on theoretical performance, neglecting practical aspects. 
In CF-MMIMO, involving more and more APs and users in the joint coherent processing leads to the network scalability issues described in~\cite{Interdonato_Scalability2019,Bjornson_Sanguinetti_Scalable_TCOM2020}. 
With the terminology ``network scalability'', we generally mean the ability of the network to handle a growing load of work as the number of APs and users increases, while fulfilling the quality-of-service (QoS) requirements. 

A main solution to preserve the scalability of the network consists in decentralizing as much as possible the network tasks and confining the coherent processing within a handful of properly selected and adjacent APs. These are the basic principles behind the \textit{dynamic cooperation clustering} (DCC) framework advocated in~\cite{Bjornson2013d}, which revisits the overlapped clustering architecture proposed in~\cite{Caire2010}, and consists in setting up tailored clusters of nearby APs, upon the knowledge of the channel statistics, to surround any user in the network. This ensures to remove the inter-cluster interference compared to a static clustering since a moving user causes clustering reconfiguration and thereby is more unlikely to stand at the edge of a cluster. Moreover, DCC confines the cooperation within a handful of adjacent APs, thereby alleviating the fronthauling load. Many valid strategies have been investigated to implement DCC such as channel statistics-dependent~\cite{Hien:TGCN:2018,Buzzi2019c,Attarifar2020,Riera2018} and received power-dependent AP selection schemes~\cite{Hien:TGCN:2018}, and the joint pilot assignment and AP selection scheme proposed in~\cite{demir2021foundations}. The common objective of these strategies is to select the best (possibly overlapped) subset of APs to serve each user so as to confine the cell-free processing within a cluster without significant loss in performance.

On a parallel research track, significant efforts have been spent to find an optimal trade-off between the performance and degree of cooperation among the APs. A network-wide fully centralized operation is optimal from a performance viewpoint but hard to implement in practice. Conversely,  in most cases, a fully distributed operation where most of the network tasks are carried out locally has low implementation requirements but performs modestly.  Some papers advocated the use of local processing (e.g. local MR or local MMSE)  at each AP  \cite{Hien:cellfree,Nayebi2017,Femenias2020_ncb,Interdonato2021,bashar2021limited}, while better performance can be achieved at the expense of higher complexity by using partially or fully centralized processing~\cite{Interdonato_PZF_2020,Attarifar_WCL2019,Liu_Luo_TWC_2019,Bjornson_Sanguinetti_TWC2020,Du2021}.

\section{Ultra-Dense Networks: State of the Art and Key Challenges} \label{Sec:UDN_SoA}

CF-MMIMO offers the potential to very significantly increase the data rate and reliability for all users in the networks.  To achieve this potential, the network must be quite dense, such that several APs are within the coverage of the typical user.  Without being able to communicate with multiple APs, the cell-free paradigm is impossible to achieve.  Thus, for CF-MMIMO, we are interested in the ultra-dense regime of network deployment, which can be defined in various ways \cite{LopezDingBook,Gotsis16,Liu16,NugKou17}, but in essence means that each location in the network is covered by multiple APs, and that, without AP cooperation, the network is strongly interference-limited as opposed to noise-limited.   More quantitatively, in the ultra-dense regime, $\lim_{M \to \infty} \sinr = \sir$, or equivalently, the ratio of the noise power to the cumulative interference power is (approximately) 0, where $M$ is the number of APs \cite{GupZha15}.   

One important benefit of CF-MMIMO is its ability to reduce the cumulative interference power through highly directional beamforming and cooperation amongst the APs.  Without any  effort to reduce interference, in many deployment and propagation scenarios, UDN networks can suffer from a ``densification collapse'', meaning that both the SINR and even the sum throughput saturate and then tend to zero after a certain level of densification is reached \cite{ZhaAnd15, AhmadUnified19,AndZha16}.  In this sense, CF-MMIMO is a method for escaping the ``densification collapse'' that occurs in many UDN deployment scenarios if interference is not ameliorated with antenna directionality \cite{Bai2014, AlmAnd20-Asilomar} or other sophisticated forms of interference reduction such as BS cooperation \cite{CloudRAN}.  

\subsection{Practical Challenges to Densification}

Achieving high network densification has proven quite challenging in practice.  Regardless of any concerns about densification collapse, the reality is that the overwhelming majority of cellular network deployments are far from the UDN regime defined above and remain coverage-limited. Globally, cellular network performance  would benefit greatly from increased densification, with the benefit arising just from standard cell-splitting gains and improved coverage, since most deployments are far from the UDN regime.   

What is holding back higher network densification? The main factor is unfortunately cost, as opposed to technical challenges such as challenges arising from increased interference levels.  Despite significant enthusiasm in the early 2010s about small-cell deployments as typified by \cite{Bhushan14,AndCla12,GhoAnd12}, the main obstacle has been that small-cell site acquisition and deployment has not proven much cheaper on a per site basis than macrocell deployments, despite the small-cell APs being much less expensive and easier in principle to deploy and configure.  This is because three of the major costs to AP deployment are site rental, legal obstacles and paperwork such as permitting, and backhaul (and power) provisioning.   These three costs vary from market-to-market, but are not significantly less expensive for small-cells in many cases.  Although directional beamforming via mmWave technology holds out the promise of eliminating the necessity of fiber connections to each cell, for example through the Integrated Access and Backhaul (IAB) feature introduced in the second release of 5G \cite{CudGho21}, the other challenges remain. 

In some cities and countries, the legal and permitting issues mentioned above are difficult to overcome irrespective of cost, for example due to strict laws in some countries regarding the total amount of electromagnetic radiation allowed \cite{EmRad18}.  This is irrespective of the science of radiation, since on aggregate for a given coverage area, a small-cell network typically consumes and radiates considerably less total power than a macrocell network \cite{EMTut15}.   There have been numerous and sometimes successful lobbying efforts by the wireless industry to lower the regulatory barriers to small-cell deployments, but this remains a major hurdle as we head towards 6G.   National governments that wish for their countries to assume leadership in 6G deployments should undertake  concerted efforts to lower regulatory burdens -- many of which are artificial and not based on science or the public interest -- to dense cellular network deployments. 

\subsection{Densification is Essential for 6G: The Road Forward}

Putting aside the economic and regulatory barriers to UDN deployments, it is important to emphasize that the news for UDNs is not all bad.  As we move towards 6G, both the appeal and viability of high network densification will continue to increase \cite{series2023imt}.   In addition to high density being a prerequisite for utilizing higher carrier frequencies including mmWave and Sub-THz, other emerging 6G features like joint communication and sensing (JCAS), including radar, require high density for accurate sensing.  Radar and visible light sensing usually require line-of-sight, which inherently means a UDN network.  Such sensing features will be indispensable for applications such as extended reality (XR) and driverless cars.   

Energy consumption is a major focus for 6G.  Perhaps paradoxically, as mentioned above, higher network density can also reduce overall radiated power due to the path loss usually being more severe than an inverse square law (whereas AP density follows units of area, which is the square of distance). Furthermore, densification enables aggregation of computing resources to the edge-cloud, and thus each individual AP becomes a very simple lower power radio unit (RU), that performs only lower physical layer functions.  This is the essence of the disaggregated RAN, typified by O-RAN \cite{polese2023understanding,polese2023empowering}.   Once the network is densified, it stays densified.  Thus AP density is a monotonically increasing function with time, so it is a question of when, rather than if, a portion of the cellular network is ready for CF-MMIMO.  The utility of CF-MMIMO can therefore also be expected to become more relevant and useful with each passing year.

\subsection{Ultra-Dense Network Meets Cell-Free Massive MIMO: Key Challenges}

As discussed earlier, the ongoing trend of network densification is anticipated to continue. To handle the densification collapse, APs should collaborate at some levels. Thus, CF-MMIMO is expected to be a useful and practical system for UDN. It can be regarded as an UDN incorporating AP cooperation with user-centric manner and leveraging massive MIMO-based techniques. Network densification introduces numerous challenges for CF-MMIMO, including the development of practical and scalable system architectures, signal processing (i.e. UL receivers and DL precoders), fronthaul designs (i.e. data routing and quantization), channel state information acquisition (i.e. channel covariance matrix and pilot contamination), and resource allocation (i.e. power control and user scheduling). Furthermore, synchronization and calibration issues come to the forefront. These challenges will be thoroughly discussed in the next sections.

\section{User-Centric Cell-Free Massive MIMO System Model}\label{sec:systemmodel}

We introduce a general mathematical model for user-centric CF-MMIMO, which will be used later in Section~\ref{sec:questions}. The system comprises $M$ APs and $K$ users distributed within a defined coverage area. Each AP is equipped with $N$ antennas, while each user has a single antenna. 
A user-centric approach is deployed, meaning that each user is served  by a preferred set of APs selected based on its needs. Denote by $\mathcal{A}_k$ the set of APs serving user $k$, and correspondingly, by $\mathcal{U}_m$  the set of users served by AP $m$. The canonical\footnote{With the terminology ``canonical'' CF-MMIMO, we mean a system where all  APs collaborate in serving all users by leveraging a single CPU through fronthaul links. This canonical configuration, introduced in \cite{Hien:cellfree}, serves as an idealized model. More details are given in Section~\ref{subsec:ORAN}.} CF-MMIMO and the small-cell systems \cite{Hien:cellfree} are the special cases of user-centric massive MIMO when $\left|\mathcal{A}_k\right| = M$, and $\left|\mathcal{A}_k\right| = 1$ for all $k$, respectively.

The channel between the $m$-th AP and the $k$-th user is an $N\times 1$ vector, denoted by
\begin{align}\label{eq:gmk}
\B{g}_{mk} = \beta_{mk}^{1/2}\B{h}_{mk},
\end{align}
where $\beta_{mk}$ represents the large-scale fading, and $\B{h}_{mk}$ is an $N\times 1$ vector of small-scale fading coefficients. We assume that the elements of $\B{h}_{mk}$ are normalized such that $\E\{|{h}_{mk,n}|^2\}=1$, where ${h}_{mk,n}$ is the $n$-th element of $\B{h}_{mk}$.  Note that the channel given by (8) can be of any type. In an extreme case, it can be LoS, i.e., $\E\{|{h}_{mk,n}|^2\}=|{h}_{mk,n}|^2=1$. However, in the following, we consider more practical scenarios where the channel includes both LoS and non-LoS components.

As in co-located massive MIMO, TDD is used in CF-MMIMO. Under TDD operation, each coherence interval of $\tau_c$ symbols includes three main activities: UL channel estimation ($\taup$ symbols), UL payload data transmission  ($\tauu$ symbols), and DL payload data transmission ($\taud$ symbols). Here, $\tauc=\taup+\tauu+\taud$.

\subsection{Uplink Channel Estimation}\label{sec:ULCE}

In the UL channel estimation phase, all $K$ users simultaneously transmit their pilot sequences to all APs. Let $\sqrt{\taup}\pmb{\varphi}_k \in
\mathbb{C}^{\taup\times 1}$, where $\|\pmb{\varphi}_k\|^2=1$, be
the pilot sequence of user $k$. Then, the pilot signal received at the $m$-th AP is 
\begin{align}\label{eq:pilot1}
	\B{Y}_{\mathrm{p},m}  
	= 
	\sqrt{\taup
	\Pp}\sum_{k=1}^K \B{g}_{mk} \pmb{\varphi}_k^H +
	\B{N}_{\p,m},
\end{align}
where $\Pp$ is the normalized transmit SNR of each pilot symbol, and $\B{N}_{\p,m}$ is an $N\times \taup$ noise matrix whose elements are
independent and identically distributed (i.i.d.) $\CG{0}{1}$ random variables.

 The received pilot matrix $\B{Y}_{\mathrm{p},m}$ will be used to estimate all channels from AP $m$ to users it serves. In general, this can be done locally at each AP or at the CPU depending
the specific processing levels. If the channels are estimated at the CPU, the APs need to forward their received pilot signals to the CPU via fronthaul links. Note that, since the channels vectors are assumed to be independent, there is no difference in the channel estimates whether the estimation is carried at the APs or the CPU. Thus, in the following, we consider the case where channels are estimated locally at each AP.
To estimate the channels $\B{g}_{mk}$, $k=1, \ldots, K$, the $m$-th AP first performs a de-spreading operation by multiplying the received signal by the pilot sequence $\pmb{\varphi}_k$,
\begin{align}\label{eq:tidley1}
\check{\B{y}}_{\mathrm{p},mk} 
	&\triangleq
\B{Y}_{\mathrm{p},m} \pmb{\varphi}_k \nonumber\\
	&= \sqrt{\taup
	\Pp} \B{g}_{mk}  + \sqrt{\taup
	\Pp}\sum_{k'\neq k}^K \B{g}_{mk'} \pmb{\varphi}_{k'}^H \pmb{\varphi}_k +
	\tilde{\B{n}}_{\p,mk},
\end{align}
where $\tilde{\B{n}}_{\p,mk}\triangleq \B{N}_{\p,m}\pmb{\varphi}_k$. Since $\|\pmb{\varphi}_k\|^2=1$, $\tilde{\B{n}}_{\p,mk}$ has i.i.d.\ $\CG{0}{1}$ components. Then, the $m$-th AP adopts the linear minimum mean-square error  (MMSE) method to estimate the channel as
\begin{align}\label{eq:MMSE est1}
&\hat{\B{g}}_{mk} 
	= \EX{\B{g}_{mk}} + \nonumber\\
	&\cov{\B{g}_{mk}, \check{\B{y}}_{\p,mk} } \cov{\check{\B{y}}_{\p,mk}, \check{\B{y}}_{\p,mk} } ^{-1}\! \!\left( \check{\B{y}}_{\p,mk}-  \EX{\check{\B{y}}_{\p,mk}}\right).
\end{align}
 
Denote by $\tilde{\B{g}}_{mk} ={\B{g}}_{mk} -\hat{\B{g}}_{mk}$ the channel estimation error. From the properties of linear MMSE estimation, $\tilde{\B{g}}_{mk}$ and $\hat{\B{g}}_{mk}$ are uncorrelated. 

\subsection{Uplink Payload Data Transmission}\label{sec:UL}
In this phase, all $K$ users send their data symbols to the APs. Denote by $\sqrt{\eta_{k}^\ul}s_k$ the  signal sent by user $k$, where $s_k$, with $\EX{|s_k|^2}=1$, is the data symbol, and $\eta_{k}^\ul$ is the UL transmit power. Then the $N\times 1$ signal vector received at AP $m$ can be expressed as
\begin{align}\label{eq:ul1}
	\B{y}_{m}^\ul  
	= 
	\sum_{k=1}^K  \sqrt{\eta_{k}^\ul} \B{g}_{mk}s_k +
	\B{n}_{m}^\ul,
\end{align}
where $\B{n}_{m}^\ul\sim \CG{\B{0}}{\B{I}_N}$ is the noise vector.

To detect the symbol sent by user $k$, all signals received by APs that serve user $k$ will be combined using the channel estimates acquired in the channel estimation phase. To this end, the combined signal used for detecting $s_k$ is given by
\begin{align}\label{eq:ul1:received}
	r_{k}^\ul  
	&= 
	\sum_{m\in \mathcal{A}_k}  \left(\B{w}_{mk}^{\ul}\right)^H \B{y}_{m}^\ul\nonumber\\
	&= \sum_{m\in \mathcal{A}_k} \sqrt{\eta_{k}^\ul} \left(\B{w}_{mk}^{\ul}\right)^H  \B{g}_{mk}s_k \nonumber\\
	&+\!\! \sum_{k'=1,k'\neq k}^K \sum_{m\in \mathcal{A}_k}\!\! \sqrt{\eta_{k'}^\ul} \left(\B{w}_{mk}^{\ul}\right)^H\!\!  \B{g}_{mk'}s_{k'}
	\!+\! \sum_{m\in \mathcal{A}_k} \!\! \left(\B{w}_{mk}^{\ul}\right)^H \!\!\B{n}_{m}^\ul,
\end{align}
where $\B{w}_{mk}^{\ul}$ is the combining vector, which is a function of the channel estimates. Note that the design of $\B{w}_{mk}^{\ul}$ depends the specific processing levels involved. For example,  in a distributed processing setup, $\B{w}_{mk}^{\ul}$ can be calculated locally at AP $m$ using only local channel estimates (i.e. the estimates of channels between itself and the different UEs). By contrast, in centralized systems, $\B{w}_{mk}^{\ul}$ can be computed at a central processing unit (CPU) connected to  APs serving user $k$, using the global channels estimates, i.e. the estimates of channels from all APs to all users.

\subsection{Downlink Payload Data Transmission}\label{sec:DL}

In the DL, for each AP, the channel estimates obtained in the UL training phase are used to precode symbols intended to only users that AP is serving.  Denote by $q_k$, where $\EX{|q_k|^2}=1$, the symbol intended for the $k$-th user. Then transmitted signal vector at the $m$-th AP is given by
\begin{align}\label{eq:xm}
\B{x}_m = \sum_{k\in \mathcal{U}_m}
\sqrt{\eta_{mk}^\dl}\B{w}_{mk}^\dl q_k,
\end{align}
where $\B{w}_{mk}^\dl$ is the corresponding precoding vector that depends on the channel estimates,\footnote{
Similarly to the UL case, the design of $\B{w}_{mk}^\dl$ depends on specific processing levels. For example, in distributed systems, $\B{w}_{mk}^\dl$ can be computed at each AP $m$ using local channel estimates. In contrast, in centralized systems, $\B{w}_{mk}^\dl$ can be computed at the CPU using global channel estimates.
} and $\eta_{mk}^\dl$ is the power control coefficient chosen to satisfy the power constraint $\Pd$ at each AP as
\begin{align}\label{eq:xm1}
\EX{\|\B{x}_m\|^2} = \sum_{k\in \mathcal{U}_m}
\eta_{mk}^\dl\EX{\|\B{w}_{mk}^\dl\|^2 }\leq \Pd.
\end{align}

With the transmitted signal vector $\B{x}_m$ given in \eqref{eq:xm}, the signal received at the $k$-th user is
\begin{align}\label{eq:rk_dl1}
r_{k}^\dl 
	&= 
	\sum_{m=1}^M \B{g}_{mk}^H \B{x}_m +  n_{k}^\dl \nonumber\\
	&=
	\sum_{m\in\mathcal{A}_k} \sqrt{\eta_{mk}^\dl} \B{g}_{mk}^H \B{w}_{mk}^\dl q_k \nonumber\\
	&+ \sum_{k'=1, k'\neq k}^K\sum_{m\in\mathcal{U}_{k'}} \sqrt{\eta_{mk'}^\dl} \B{g}_{mk}^H \B{w}_{mk'}^\dl q_{k'} +  n_{k}^\dl,
\end{align}
where $ n_{k}^\dl \sim \CG{0}{1}$ is  additive noise.

\section{Key Open Questions} \label{sec:questions}

\subsection{Will Cell-Free Massive MIMO Fit Well into O-RAN Architectures? }
\label{subsec:ORAN}

Designing a scalable and practical architecture for CF-MMIMO is a major challenge. 
%
%
As mentioned earlier, from a theoretical point of view, the user-centric approach is a scalable method for implementing CF-MMIMO. Forming user-centric clusters can be done via several simple methods: for example,  each user can choose some of its closest APs, or opt for a subset of APs that collectively contribute the highest received desired signal power  \cite{Hien:TGCN:2018}. More complicated methods, such as joint user association and power control, were exploited in \cite{hao2023user}.  However, the main challenge in user-centric CF-MMIMO is how to control the user-centric clusters. Conventionally, each cluster needs to be controlled by one or several CPUs. But these clusters change quickly with user load and locations. This requires more control signaling and poses a huge challenge. In \cite{Interdonato_Scalability2019,riera2019decentralization}, a hybrid approach was studied, where the CF-MMIMO network was shaped by several network-centric clusters, while each user forms its own user-centric  cluster. This approach is still fully network-controlled, and still requires huge connection bandwidth from all APs to the CPUs, especially when the network size increases.  So the key question is: what is a practical architecture to support user-centric CF-MMIMO?

Recently, \textit{Open Radio Access Network} (O-RAN) has emerged as a  promising approach for the organization of mobile networks \cite{polese2023understanding,polese2023empowering,abdalla2022toward}. This innovative concept is fundamentally reshaping the radio access network (RAN) landscape, transforming it toward an open, intelligent, virtualized, and fully interoperable architectural framework. With O-RAN, we are witnessing the unlocking of internal RAN interfaces, which, in turn, is fostering the creation of an open ecosystem where multiple vendors can contribute diverse solutions, resulting in a complete, cost-efficient end-to-end network ecosystem. The open interfaces facilitate the integration of cutting-edge technologies, enhancing the performance of mobile communication systems. Meanwhile, the utilization of software-enabled implementations for the baseband processing unit  accelerates the pace of development and facilitates seamless upgrades. The architecture and features of O-RAN provides an excellent platform for the practical implementation of cell-free systems \cite{ranjbar2022cell}. The main reasons are:
\begin{itemize}
\item In O-RAN, the physical layer (PHY) architecture is divided into two key components: the O-RAN Radio Units (O-RUs) manage low PHY operations, while the O-RAN Distributed Units (O-DUs) handle high PHY functions \cite{garcia2021ran,huang2023validation}. This design closely resembles the user-centric CF-MMIMO system model, where PHY processing is  divided between between the APs and the CPUs \cite{ranjbar2022cell}.  Note that, depending on the context and specific requirements of the operator, O-DU and O-RU can either be co-located or deployed separately \cite{garcia2021ran}.

\item Additional  functional blocks have been incorporated to enhance the capabilities of the network, facilitating AI integration and containerized service orchestration.  These critical additions include the Near-Real Time RAN Intelligent Controller (NearRT RIC), the Non-RT RIC, and the robust Service Management
and Orchestration (SMO) framework \cite{polese2023understanding}. The functional split
 along with expanded options for network-wide control can achieve cooperation amongst O-RUs, even beyond the borders of the O-DUs.

\item O-RAN offers an advanced clock and time synchronization mechanism for multiple RUs. With over-the-air (OTA) reciprocity calibration \cite{rogalin2014scalable}, the DL transmission has been successfully tested and validated within experimental systems utilizing commercially available off-the-shelf O-RUs \cite{cao2023experimental}.

\end{itemize}

\begin{figure}[t!]
\centering
\subfigure[User-centric cell-free massive MIMO.]{%
\includegraphics[width=0.40\textwidth]{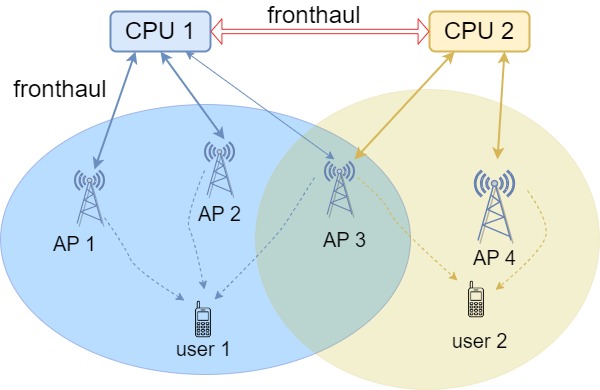}
\label{fig:UCCFMM}} \quad
\subfigure[O-RAN architecture for cell-free massive MIMO.]{%
\includegraphics[width=0.40\textwidth]{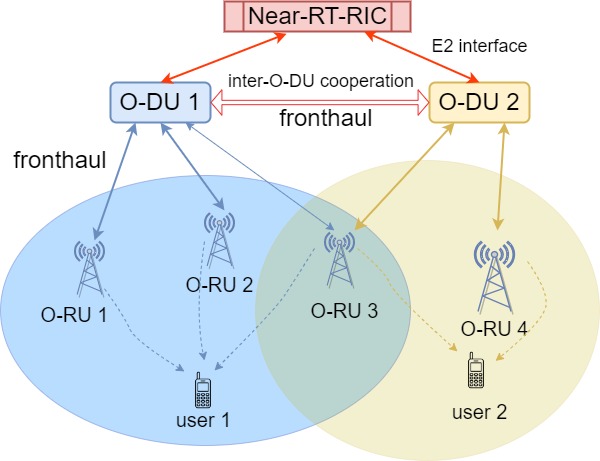}
\label{fig:UCCFMM_ORAN}} \caption{Cell-free massive MIMO fits well into the O-RAN paradigm.} \label{fig:CFORAN}
\end{figure}

Therefore, the O-RAN architecture is a suitable match for cell-free networks, possibly facilitating  the practical realization of future CF-MMIMO configurations  \cite{ranjbar2022cell,demir2024cell,oh2023decentralized}. Figure~\ref{fig:UCCFMM} represents theoretical user-centric CF-MMIMO.  Figure~\ref{fig:UCCFMM_ORAN} represents a corresponding (possible) practical implementation leveraging the O-RAN architecture, where the CPUs are represented by the O-DUs, and the APs are represented by
O-RUs. Multiple O-DUs are connected to the near-real-time RAN intelligent controller (Near-RT RIC)  through the E2 interface for centralized network performance control, such as user categorization and management. Furthermore, within the O-RAN framework, the Physical Layer (PHY) functions are distributed between the O-DU and O-RU following 7-2x option \cite{ranjbar2022cell}. More precisely, O-DU performs (de-)modulation, channel estimation, and equalization, while O-RU performs digital beamforming. 
Notably, the O-RAN split 7-2x encompasses two distinct categories of O-RUs: category A and category B. Category B O-RUs possess the capability to execute precoding operations, while Category A O-RUs lack this functionality.

In CF-MMIMO, each user should be served by O-RUs that are connected to different O-DUs to effectively mitigate the interference, and hence, improve the system performance. This requires cooperation among these O-DUs. The inter-O-DU cooperation interface can be used to exchange user data between O-DUs and to send necessary signaling required for the computation of the combining and precoding  vectors in the UL  and DL  transmission. Thus, inter-O-DU coordination is critical for enabling the full potential of a CF-MMIMO network deployment. However,  the current O-RAN standard lacks a well-defined framework for the inter-O-DU coordination interface. This underscores the need for a rigorous specification of the signaling mechanisms required for inter-O-DU cooperation. Developing this signaling framework is a complex task that requires further investigation. In addition, to ensure scalability with the system architecture, adopting a \textit{flexible fronthaul} and incorporating cluster processors as software-defined network functions capable of dynamic migration alongside users throughout the network, emerges as a promising solution. However, it is worth noting that the  implementation of this approach poses significant challenges and calls for a comprehensive exploration of potential solutions.

\subsection{What User-Centric  Processing Should Ultra-Dense Cell-Free Massive MIMO Use? }

In Section~\ref{sec:systemmodel}, we  described a general user-centric CF-MMIMO system where the signal processing, including UL combining and DL precoding, can be done centrally at the CPUs or locally at each AP. There are two  main types of processing:
\begin{itemize}
\item Centralized signal processing: all APs in a cluster are connected to a CPU. The UL combining and DL precoding vectors (i.e., $\B{w}_{mk}^\ul$ and $\B{w}_{mk}^\dl$) are designed at the CPU. The optimal processing is MMSE combining/precoding \cite{demir2021foundations}. The centralized signal processing requires the exchange of pilot signals, received data signals, data symbols, and precoding vectors between the CPU and the APs via fronthaul links. In addition, tasks such as channel estimation, precoding computation, and combining computation are executed at the CPUs, resulting in substantial computational complexity, particularly in UDNs where there is a concentration of many APs and users in a small space. Note that the DL transmission scheme described above can be viewed as the compression-before-precoding scheme in CRAN. Alternatively, the precoded signals can be computed at the CPU first and then sent to the APs. This approach is similar to compression-after-precoding in CRAN \cite{kang2015fronthaul}. 

\item Local signal processing: the channel estimates,  UL combining and DL precoding vectors are computed locally at each AP. In the UL, only the scalars representing the local detected versions of received signals are transmitted from the APs to the CPUs for the cluster-level signal detection. In the DL, only data symbols need to be sent from CPUs to the APs. 
\end{itemize}

\begin{figure}[t!]
\centering
\subfigure[Uplink.]{%
\includegraphics[width=0.44\textwidth]{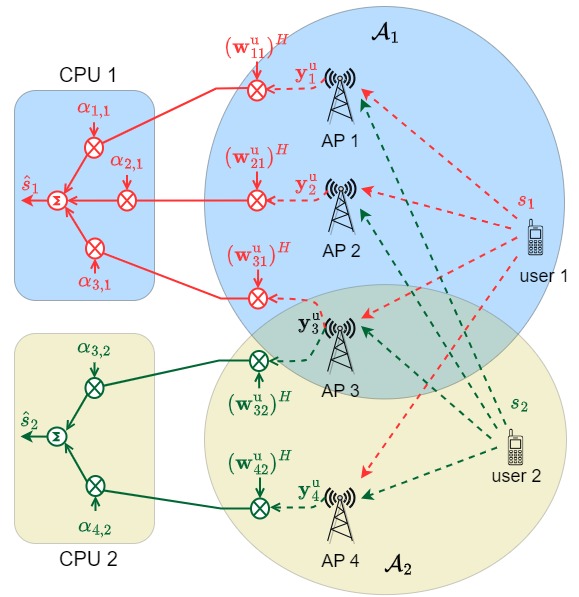}
\label{fig:LocalCFMM_UL}} \quad
\subfigure[Downlink.]{%
\includegraphics[width=0.44\textwidth]{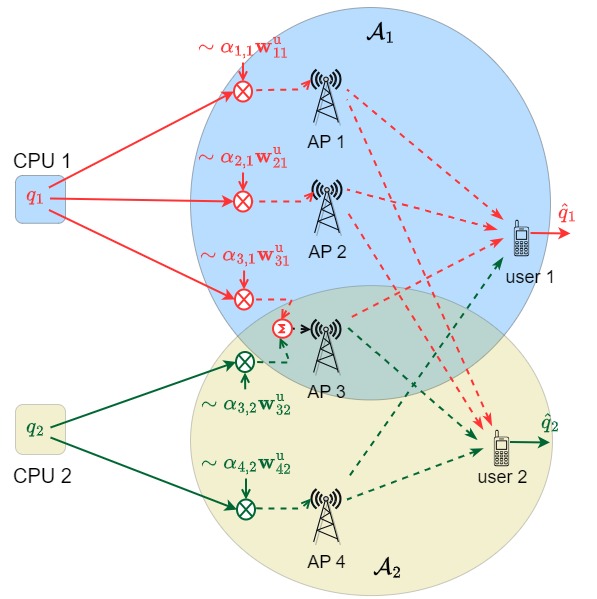}
\label{fig:LocalCFMM_DL}} \caption{User-centric cell-free massive MIMO with local processing.  Here we have 4 APs and 2 users.  User 1 is served by APs 1, 2, and 3, while user 2 is served by APs 3 and 4. This means $\mathcal{A}_1 = \{1, 2, 3\}$, $\mathcal{A}_2 = \{3, 4\}$, $\mathcal{U}_1 = \{1\}$, $\mathcal{U}_2 = \{1\}$, $\mathcal{U}_3 = \{1, 2\}$, and $\mathcal{U}_4 = \{2\}$.} \label{fig:LocalCFMM}
\end{figure}

It is foreseen that local processing is suitable  for ultra-dense CF-MMIMO, because (i) the channel estimation and the computation of precoding/combining vectors are performed at each AP. This significantly reduces the computational complexity at the CPUs; 
(ii) the system is robust in the sense that the addition or failure of some APs does not affect the overall system design; and (iii) it performs well with multiple-antenna APs \cite{gottsch2022uplink}. Thus, in the following, we will focus on the implementation of the local processing. All notation will be the same as in Section~\ref{sec:systemmodel}.

\subsubsection{Uplink Payload Data Transmission}

To detect the signal transmitted by user $k$, all APs serving user $k$ will first locally estimate $s_k$ from their received signals. More precisely, AP $m$, where $m\in \mathcal{A}_k$, produces a local detection of symbol transmitted by user $k$, in the form
\begin{align}\label{eq:ul_process_1}
	r_{mk}^\ul  
	&= 
	\left(\B{w}_{mk}^{\ul}\right)^H \B{y}_{m}^\ul,
\end{align}
where $\B{y}_{m}^\ul$ is the received signal vector at AP $m$, given by \eqref{eq:ul1}, and $\B{w}_{mk}^{\ul}$ is an $N\times 1$ local detection vector that can be computed locally at AP $m$ using its channel estimates acquired during the UL channel estimation phase. One example of the local detector is MR combining, for which $\B{w}_{mk}^{\ul}= \hat{\B{g}}_{mk}$ \cite{bashar2018mixed}. Other techniques (which have better performance but higher computational complexity) are the local ZF, local MMSE \cite{demir2021foundations,gottsch2022subspace}, and  the \textit{team} MMSE (TMMSE) method ~\cite{zheng2022team,Miretti2021_team,miretti2022team}.

Then the $m$-th AP sends its detected version $r_{mk}^\ul$ to the CPU hosting 
the cluster processor for user $k$. This CPU receives $|\mathcal{A}_k|$ signals $r_{mk}^\ul$ for all $m\in \mathcal{A}_k$, and produces a cluster-level combined signal as
\begin{align}\label{eq:ul_process_2}
	r_{k}^\ul  
	&= 
	\sum_{m\in\mathcal{A}_k}\alpha_{mk} r_{mk}^\ul,
\end{align} 
where $\{\alpha_{mk}\}$ are the combining coefficients chosen to manage the interuser interference and noise. Intuitively, if $r_{mk}^\ul$ is poor (i.e. it is interference-and-noise dominated), then AP $m$ should send its small version or  not even send it the CPU, i.e. $\alpha_{mk}$ should be relatively small. By contrast, when $r_{mk}^\ul$ is good, it is advantageous to allocate a larger value to $\alpha_{mk}$. The optimal values of 
$\{\alpha_{mk}\}$ can be efficiently computed by solving a generalized eigenvalue problem to maximize the receive SINR as in \cite{nayebi2016performance,Bashar:TWC:2019}. Alternatively, they can be chosen to maximize the ``nominal'' SINR where the out-of-cluster interference is replaced by a corresponding white noise term  \cite{gottsch2022subspace}. This approach addresses the practical scenarios where a given CPU knows only the  channel statistics within its cluster. In addition, depending on the designs, $\alpha_{mk}$ can be a function of the large-scale  or small-scale fading. In the later case, the instantaneous channel estimates need to be sent from APs to the CPUs for computing the optimal $\alpha_{mk}$. This approach is not exact local processing, as discussed before, but is also a suitable method for practical implementation.

Finally, the data symbol $s_k$ is detected by treating $r_k^\ul$ as a virtual single-user additive noise channel.

The above local processing in the UL can be characterized as the ``local detection with cluster-level combining''. Figure~\ref{fig:LocalCFMM_UL} demonstrates an example of this scheme.

\begin{Remark}
As mentioned above, each AP $m$ first sends its detected version $r_{mk}^\ul$ to the CPU. Then the CPU will combine these signals using \eqref{eq:ul_process_2}. This is implemented under the assumption that the fronthaul is a mesh network of arbitrary topology formed by error-free capacity links.  Alternatively, each AP $m$ can first scale its locally detected signal with the weighting factor $\alpha_{mk}$ before sending to the CPU, i.e. AP $m$ will send $\alpha_{mk} r_{mk}^\ul$ to the CPU. By doing this, all signals can be combined OTA, or the CPU can decode the sum given in \eqref{eq:ul_process_2}. This approach requires AP $m$  to know $\alpha_{mk}$. This coefficient can be computed at the CPU and sent to AP $m$, or each AP $m$ can compute $\alpha_{mk}$, given that it knows some statistical channel knowledge. 
\end{Remark}

\subsubsection{Downlink Payload Data Transmission}
In the DL, each AP $m$ first  computes the precoding vectors combined with the power control coefficients $\eta_{mk}^\dl$ as  $\sqrt{\eta_{mk}^\dl}\B{w}_{mk}^\dl$ for all $k\in \mathcal{U}_m$. Interestingly, motivated by the approximate UL-DL duality,\footnote{Note that if the hardening bounding technique is used, the UL-DL duality can be exact \cite{demir2021foundations,miretti2023ul}.}
the DL precoding vectors and power control coefficients can be directly obtained by using the UL combining vector $\B{w}_{mk}^\ul$ and cluster-level combining coefficient $\alpha_{mk}$. More precisely, $$\sqrt{\eta_{mk}^\dl}\B{w}_{mk}^\dl =  \frac{\sqrt{\Pd}}{\sqrt{|\mathcal{U}_m|\EX{\|\bar{\B{w}}_k \|^2}}}\alpha_{mk} \B{w}_{mk}^\ul,$$ where $\bar{\B{w}}_k$ is obtained by stacking $\alpha_{mk} \B{w}_{mk}^\ul$, $\forall m\in \mathcal{A}_k$, on top of each other \cite{gottsch2022subspace,li2023joint}. This means that there is no extra computation for the DL precoding if the UL and DL use the same processing scheme. An example of the local DL processing is shown in Figure~\ref{fig:LocalCFMM_DL}.

In summary, local processing stands out as an ideal choice for ultra-dense CF-MMIMO systems. It provides an excellent balance between the computational complexity,  system resilience, and system performance. The remaining challenge in implementing this approach is the necessity for a flexible fronthaul infrastructure capable of dynamically routing fronthaul traffic among dynamic clusters.

\subsection{How to Dynamically Route Fronthaul Traffic and Quantize Signals?}

As discussed in previous sections, since users are served by user-centric clusters of APs that depend on the user locations and are not a priori determined, a  requirement for scalability is that APs and CPUs are connected by a flexible fronthaul network, 
able to route the data traffic between APs and CPUs. Without loss of generality, we can assume that such flexible fronthaul is formed by point-to-point wired or wireless links\footnote{In the case of wireless links, we assume that they are implemented in a frequency band not interfering with the radio access network and make use of high-gain beamforming to eliminate inter-link interference.} 
and intermediate routing-capable nodes, referred to  as {\em routers} (see  Fig.~\ref{fig:UCCFMM}). 

Each CPU has a finite computation capacity and can host up to a given maximum number of {\em cluster processors}. 
A cluster processor is a {\em software-defined virtual network function} 
running on the general-purpose hardware of the CPU that 
implements the PHY for the corresponding user. This includes pilot-based channel estimation in the UL, 
computation of the linear receiver vector for interference mitigation and UL detection, and UL channel decoding, 
as well as computation of the DL  precoding vectors and of the precoded signals to be transmitted jointly by the cluster of APs in the DL. 

In order to reduce the fronthaul load,  the key system design aspects are:
1) optimal routing of the UL (from APs to CPUs) and DL (from CPUs to APs) data traffic such that 
the maximum link load is minimized; 2) optimal placement of the user-centric cluster processor in the CPUs, such that 
each CPU does not violate its maximum computation capacity; 3) optimization of the number of 
{\em bits per signal dimension}\footnote{A finite energy bandpass 
	signal of single-sided bandwidth $W$ and approximate duration $T$ span a signal space of dimension
	$\approx WT$ over the complex field $\CC$, for large  $WT$ \cite{gallager1968information}. 
	As usual in communication theory, taking this limit with equality, we identify bit per signal dimensions, i.e., per complex baseband sample, 
	with the SE in bit/s/Hz.}
to represent the received signal in the UL (from APs to CPUs) and the 
user information messages in the DL (from CPUs to APs). 

For a given network topology, the load balancing (via routing) and the allocation of computation resources must be jointly optimized. 
For example, if the cluster processor for a certain user $k$ is allocated to a geographically remote 
CPU, the data between the CPU and the APs forming the cluster of user $k$ must travel across many hops in the fronthaul 
network, thus generating a higher overall load. On the other hand, it may be impossible to allocate 
cluster processors to CPUs at a minimum number of hops, because some CPU may violate their computation capacity constraint. 

This joint optimization problem is formulated and addressed in \cite{li2023joint} in the form a of mixed-integer linear program (MILPs), 
which can be efficiently solved even for fairly large networks. Also in \cite{li2023joint} it is shown that in the UL the fronthaul routing problem is a multiple unicast problem where data sources are the (quantized) projected received signals 
	$r_{mk}^\ul= \left(\B{w}_{mk}^{\ul}\right)^H \B{y}_{m}^\ul$ 
	given by \eqref{eq:ul_process_1}, that are produced (at rate of 1 sample per s$\times$Hz) by AP $m$ in relation to its connected user $k$. For all AP $m \in \mathcal{A}_k$ forming the cluster of user $k$, these 
 ``sources''  must be routed to the CPU hosting user $k$ cluster processor, in order to form the cluster-level combining samples
 $r_k^{\rm u}$ as in \eqref{eq:ul_process_2}. In order to optimize the UL fronthaul load, a critical aspect is the fronthaul quantization, i.e., 
 the complex samples $r^{\rm u}_{m,k}$ are quantized with a finite number of bits. Using sample-per-sample dithered scalar quantization, \cite{li2023joint} identifies in the quantization squared  distortion $D$ the key parameter for fronthaul quantization optimization. Fixing $D$, the resulting 
 quantization rate can be made close to the rate-distortion limit
 $R_{mk} = \max\{ \log_2 (\sigma_{mk}^2/D), 0\}$ bits per sample, where $\sigma_{mk}^2 = {\rm Var}(r^{\rm u}_{mk})$. 
 This means that the number of quantization bits is not fixed for all observations; instead, it depends on the ratio $\sigma_{mk}^2/D$. In particular, if $\sigma_{mk}^2 \leq D$, the observation is discarded (i.e., allocated zero rate). In this way, only the ``strong'' observations are effectively quantized and forwarded to the cluster processor for cluster-level combining. 
 
In contrast, in order to minimize the fronthaul rate in the DL it is always convenient to forward the information bits rather than 
the precoded modulated (complex-valued) data streams, and implement encoding/modulation and multiplication by the precoding coefficients in the APs. In fact, the channel-encoded and modulated OFDM symbols forming a user data stream
at DL rate of $R_k^{\rm d}$ bits/symbol are best compressed by sending exactly $R_k^{\rm d}$ bit/symbol, i.e., the
(incompressible) user information bits. Note in passing, in the context of O-RAN discussed in Section~\ref{subsec:ORAN}, that this shows that the so-called 7.2 ``split'', i.e., the interface between CPU and AP, that specifies to send modulated I and Q samples in the frequency domain, is not efficient for the DL. 
Rather, ``smarter'' APs that can accomplish the coding/modulation and precoding operations are 
better suited for such network architectures.\footnote{This does not mean that the APs must  locally compute the precoding coefficients. These can be computed centrally by the CPU, and sent through the CPU-AP interface at low speed since one set of coefficients is needed for each  DL channel coherence block.}
It is also interesting to notice that in the DL, the fronthaul optimal routing problem involves multiple multicast since 
the data of each user $k$, generated at the CPU hosting the cluster processor of user $k$, must be routed
to all APs $m \in \mathcal{A}_k$, forming the user-centric cluster of user $k$. Nevertheless, the optimization framework of
\cite{li2023joint} captures also this aspect. 


\subsection{When Does the Hardening Bounding Technique Work? }
One key feature of massive MIMO is ``channel hardening'', where the effective channel gain  closely approximates its mean value \cite{ngo16,ngo2017no,emil17}. This allows accurate detection by relying only on the mean of the effective channel gain. This channel hardening-based detection  corresponds to an achievable rate obtained by using the ``hardening bounding technique'' (a.k.a. the ``use-and-then-forget bounding'' technique) \cite{ngo16,ngo2017no,emil17}, which equals the mutual information between the desired signal and the received signal, under the condition that the receiver  knows only the mean of the effective channel gain.  More precisely, the DL achievable rate based on the channel hardening bounding technique is 
\begin{align}
	R_{k}^{\CH,\dl}
	&= 
	I\left(r_k^\dl; q_k \right),\label{eq:DL_rate_hardening1}
\end{align}
where $r_k^\dl$ is  the received signal at the $k$-th user given by \eqref{eq:rk_dl1}, and $q_k$ is the date symbol intended for user $k$. A similar formula can be obtained for the UL.

The hardening bounding technique is  widely used in the massive MIMO literature \cite{ngo16,ngo2017no,emil17,van2021reconfigurable,chowdhury2022can,zheng2023rate}, for two main reasons. First, it yields  simple rate expressions that give  valuable insights. Second, in massive MIMO, under certain assumption on propagation environments (e.g. independent Rayleigh fading), the channel hardens, and hence, the hardening bounding rate is remarkable close to the full-side-information rate where the receiver uses the true effective channel gain for signal detection, given by
\begin{align}
	R_{k}^{\FSI,\dl}
	&= 
	I\left(r_k^\dl, \text{full CSI}; q_k \right).\label{eq:DL_rate_FSI}
\end{align}
In CF-MMIMO, since only a subset of APs serves a given user and the channels from different APs have different large-scale fading coefficients, the level of channel hardening may not be as pronounced \cite{chen2018channel}. As a result, the hardening-based rate  may underestimate the practical rate. Interestingly, it is shown in \cite{polegre2020new,demir2021foundations} that in most scenarios, the gap between the hardening-based rate \eqref{eq:DL_rate_hardening1} and full-side-information rate \eqref{eq:DL_rate_FSI} is relatively small, especially in  centralized setups. A substantial performance gap only arises when MR processing is used or when single-antenna APs are deployed. However, for the DL, under MR processing (a.k.a. conjugate beamforming), the above gap can be reduced by ``artificially'' hardening the channels via the use of normalized conjugate beamforming~\cite{interdonato2016performance,Femenias2020_ncb, Interdonato2021, Interdonato_uCB}. For instance,~\cite{Interdonato2021} advocates the use of the so-called \textit{enhanced conjugate beamforming} (ECB) where the precoding vector $\B{w}_{mk}^\dl \!=\! \hat{\B{g}}_{mk}/\lVert\hat{\B{g}}_{mk}\rVert^2$. How ECB is able to boost the hardening of the effective DL channel is intuitive. Let us consider the effective DL channel at the $k$-th user as
\begin{align}
    \label{eq:a_kk}
    a_{kk} &\!=\! \sum_{m\in\mathcal{A}_k} \sqrt{\eta_{mk}^\dl} \B{g}_{mk}^H \B{w}_{mk}^\dl.
\end{align}
Then, under error-free channel estimation assumption, $a_{kk}=\sum\limits_{m\in\mathcal{A}_k} \sqrt{\eta_{mk}^\dl}$, which is ideally deterministic.

The above studies assume Rayleigh fading channels, and independence between  channels from different APs, which may not hold in ultra-dense scenarios. In some practical scenarios, particularly in UDNs, channels from a given user to different APs may  not be independent,  and hence, the  channel hardening is not  guaranteed. As a result, the channel hardening bounding technique may  not be good enough. To see this, let us consider the following example.

\textbf{Example}: $M$ APs and $K$ users are uniformly located at random in an area of $1\times 1$~km$^2$. The simulation setup is similar to that of \cite{demir2021foundations}, with the following exceptions:
\begin{itemize}
\item For  small-scale fading (i.e. $\B{h}_{mk}$), we consider a simple double-scattering channel model:
\begin{align}\label{eq:double_scattering1}
    \B{h}_{mk} &=  \frac{1}{\sqrt{n}}\B{H}_{1,mk} \B{h}_{2,k},
\end{align}
where $\B{H}_{1,mk}\in \C^{N\times n}$, $\B{h}_{2,k}\in \C^{n\times 1}$, and elements of $\B{H}_{1,mk}$ and $\B{h}_{2,k}$ are i.i.d. $\CG{0}{1}$. As shown in \cite{ngo2017no}, double-scattering channels do not harden.

\item AP selection is performed based on large-scale fading, i.e., each user $k$ will choose $M_0 \leq M$ APs associated with $M_0$ largest large-scale fading coefficients $\beta_{mk}$. In addition, pilots are randomly  assigned.
\end{itemize}

\begin{figure}[t!]
\centering
\includegraphics[width=0.48\textwidth]{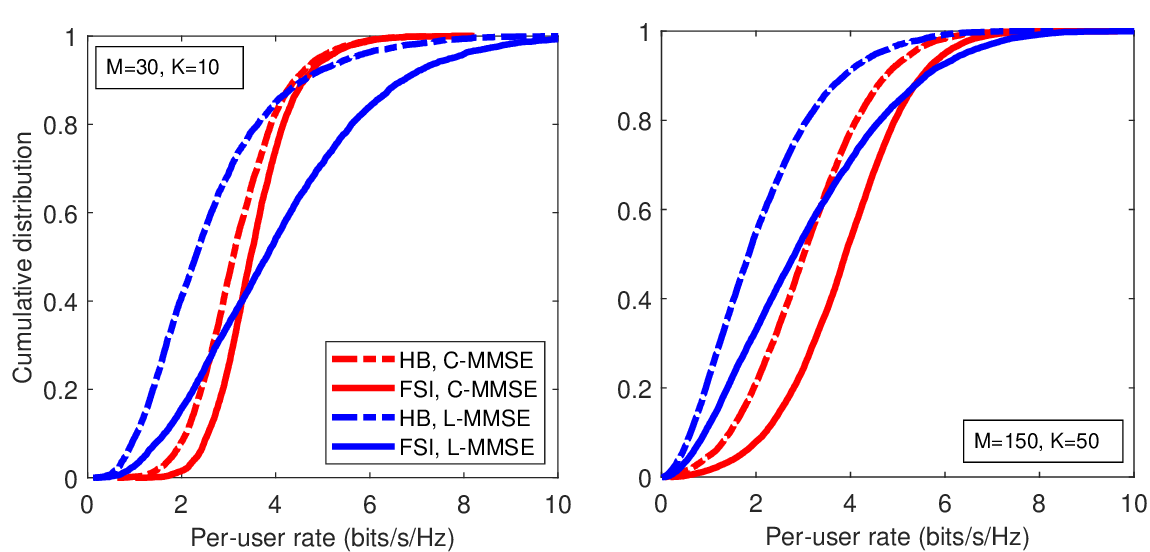}
\caption{Per-user rate under double-scattering channels. Here $N=2, M_0=10$,  $\taup =10$, and $n=1$. } \label{fig:HBVsFSI_DC}
\end{figure} 

\begin{figure}[t!]
\centering
\includegraphics[width=0.48\textwidth]{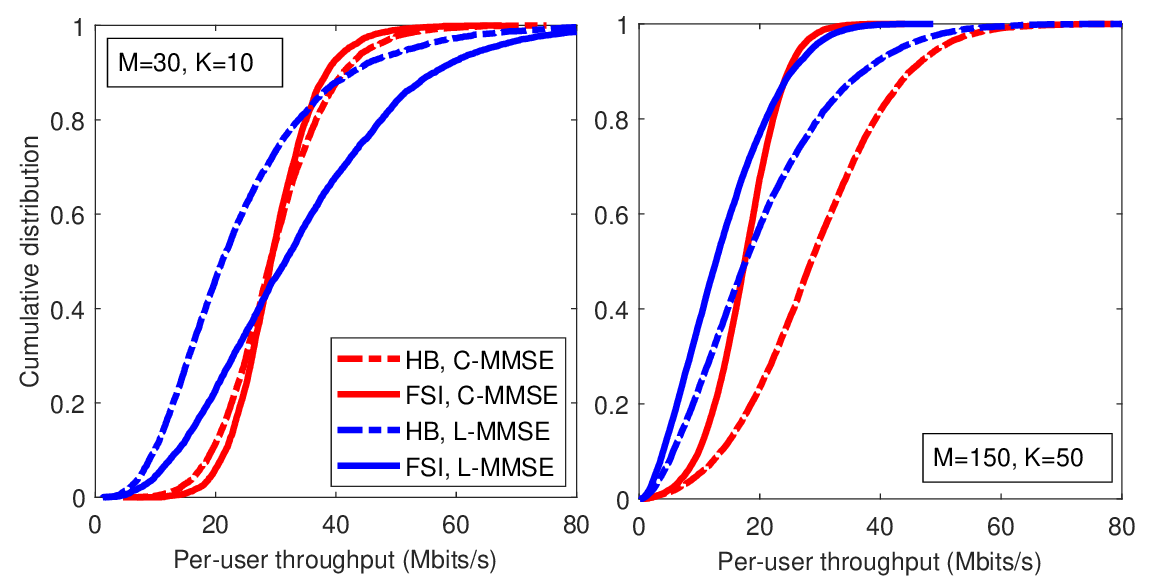}
\caption{Per-user throughput under double-scattering channels. Here $N=2, M_0=10$,  $\taup =10$, and $n=1$. } \label{fig:HBVsFSI_throughput}
\end{figure}

We consider two precoders: centralized MMSE (C-MMSE) and local MMSE (L-MMSE) \cite{demir2021foundations}. Figure~\ref{fig:HBVsFSI_DC} compares the per-user rate for hardening bounding and full-side-information bounding techniques. We can see that, in double-scattering channels, the full-side-information rate is much higher than the channel  hardening rate, especially in the dense networks (i.e. $M=150$ and $K=50$). In these scenarios, more suitable bounding techniques become necessary and the associated channel acquisition schemes should be considered. Potential approaches include:
\begin{itemize}
\item Side-information rate and DL beamforming training: to achieve the full-side-information rate, users need to have perfect knowledge of the effective channel gains. In practice, these gains need to be estimated. One possible approach is to employ a  DL beamforming training method as in \cite{ngo2013massive,interdonato2019downlink}. In this case, the DL channel estimation overhead should be taken into account, and hence, the performance of full-side-information technique may reduce significantly. To see this, let us consider the per-user throughput, defined as $S_{k}^{\CH,\dl} \triangleq \frac{\taud}{\tauc} B \times R_{k}^{\CH,\dl}$ and $S_{k}^{\FSI,\dl} \triangleq \frac{\taud-K}{\tauc} B\times R_{k}^{\FSI,\dl}$ for the hardening bounding and full-side-information techniques, respectively, where $B$ is the system bandwidth. The factor $\taud-K$ in the full-side-information formula comes from the fact that an additional duration of $K$ symbols is used for the DL training \cite{interdonato2019downlink}. Figure~\ref{fig:HBVsFSI_throughput} compares the per-user throughputs. We can see that even under double-scattering channels, if the DL channel estimation overhead is taken into account, the hardening bounding technique is even better than under double-scattering channels, especially in dense networks where we have many users (and hence, more DL pilots are required for the DL training).  Only if the number of users is small (relatively to $\taud)$ and L-MMSE is used, the hardening bound throughput falls significantly below the full-side information throughput.

\item Side-information rate and blind channel estimation: this scheme eliminates the need for DL training, thereby  preserving resources for DL channel estimation. However, the scheme 
relies on the sample average power of the received signals as well as on the law of large numbers, which works well only when the coherence interval and the number of users are large \cite{ngo2017no,souza2021effective}.

\item Non-coherent precoding-based bound: an alternative capacity lower bound that  does not rely on an explicit estimate of the instantaneous effective channel gains at the users is proposed in \cite{caire2018ergodic}. This bound remains effective even in scenarios where the channels do not harden. However, it requires that the coherence interval is  much larger the number of users. 

\end{itemize}

To summarize, the hardening bounding technique works fairly well under hardening propagation environments (i.e. independent Rayleigh channels), even when the cluster size is small or the number of users is large. Otherwise, if the propagation environments do not offer the hardening property, the variance of the effective channel gain creates strong self-interference, which limits the performance of hardening bounding technique. In these scenarios, we can use the side-information bound with DL beamforming training or a non-coherent bounding technique (if $\taud\gg K$), or side-information bound with blind channel estimation (if $\taud \gg 1$ and $K \gg 1$). We can also use enhanced normalized precoding to harden the channels \cite{sutton2021hardening,Interdonato2021}. Then the hardening bound can be used. But this scheme is applied for DL MR processing only. Alternatively, we can consider implementing some suitable scheduling techniques, like dropping users,  that reduce the channel hardening level, from service. 

Finally,  it is important to  highlight that the hardening bound is a simple lower bound on the ergodic capacity. It is not an approximation. Its main advantage is its analytical simplicity, providing a simple and insightful closed form in many scenarios, and even if not, it is just a function of the fading and noise moments. For a given value as predicted by the hardening bound one would have to design a specific code. But in practice, coding and modulation for the additive white Gaussian
noise (AWGN) channel work very well, as has been demonstrated experimentally. However, like other ergodic rate bounds, achieving this hardening bound requires the codeword to span over many small-scale fading realizations. In some practical systems, especially with bursty traffic where the users have dynamic buffers, the user scheduling is done every transmission time interval (TTI) and over narrow frequency bands. Consequently, the set of active users changes quickly over time/frequency. In such cases, this bound may not be the most suitable approach.

\subsection{How Much Prior Information is Really Available and How Much Should be Used?}

As we have seen, in CF-MMIMO it is essential that each AP obtains good-quality estimates of the channel vectors for all its associated users via UL pilots and UL/DL channel reciprocity. As shown in \eqref{eq:MMSE est1}, each user $k$ associated with AP $m$ sends an UL pilot and the estimated channel $\hv_{mk}$ is obtained by projecting the received UL pilot symbols onto the corresponding pilot sequence $\phiv_k$. Since in general the pilot dimension is less than the number of simultaneously active users $K$, the pilot sequences are generally non-orthogonal. In most system designs, and in compliance with the 3GPP 5GNR standard, 
a set of mutually orthogonal pilots~\footnote{Orthogonality can be achieved by using different symbols in the OFDM time-frequency grid.} is reused across the users in the system, and various pilot assignment schemes have been proposed such that users assigned to the same pilot are far apart in the network coverage area. 
If some user $k' \neq k$ uses the same pilot of user $k$, its contribution in the interference term in \eqref{eq:MMSE est1} does not disappear. Hence, the AP will estimate the sum of channels $\hv_{mk}$ and $\hv_{mk'}$. This fact is well-known in the massive MIMO and CF-MMIMO literature under the name of \textit{pilot contamination}  \cite{ngo16,Hien:cellfree}. 

Pilot contamination can be mitigated by exploiting the knowledge of the channel statistics. In particular, if the channel vectors exhibit  spatial correlation with directional preference, linear MMSE filtering  \cite{demir2021foundations} or the simpler channel subspace projection approach \cite{gottsch2022subspace} are able to significantly improve the UL channel estimation. In short, the estimate of the channel $\hv_{mk}$ obtained by \eqref{eq:MMSE est1} can be further improved by 
projecting it onto the dominant channel subspace, i.e., the subspace spanned by the dominant  eigenvectors of the channel covariance matrix $\Sigmam_{mk} = \cov{\hv_{mk}}$. In fact, when co-pilot users (i.e., users causing mutual pilot contamination) are well separated in the channel subspaces, the channel subspace projection approach nearly eliminates the pilot contamination. In particular, this is the case for highly directional channels, when the propagation between a user and an AP occurs along a few angular clusters, and when users are spatially distributed such that these channel subspaces are nearly mutually orthogonal in the angular domain. 

The  channel covariance matrices (or dominant channel subspaces) for each  AP-user pair $(m,k)$ must also be estimated. It is well known that the so-called channel geometry coherence time, i.e., 
the time scale at which the geometry of the multipath channels varies over time, is very large in comparison with the small-scale fading coherence time. This implies that the (vector-valued) random processes are locally wide-sense stationary (WSS) such that their second-order statistics (notably, the channel covariance matrix) is time-invariant over a large number of independent (small-scale) fading blocks. 
It is therefore commonly concluded that estimating the channel statistics is relatively easy, since this can be done across a long sequence of UL pilot intervals. Nevertheless, 
the acquisition of the channel statistics is affected by the same pilot contamination problem that its use tries to defeat. To go back to our example of above, if AP $m$ can only estimate the sum of the channel vectors $\hv_{mk} + \hv_{mk'}$ for two co-pilot users $k$ and $k'$, then also the resulting sample covariance converges to the sum of the individual channel covariances $\Sigmam_{mk} + \Sigmam_{mk'}$. 
In \cite{gottsch2022subspace}, a solution of this problem is proposed, where specially dedicated frequency-hopped UL pilot symbols are assigned geographically across the network using an assignment based on 
orthogonal latin squares of order $N$ \cite{Pottie95}. In this way, the $N$ pilot samples used to estimate the channel dominant subspace of 
$\hv_{mk}$ are affected by a very sparse pattern of strong co-pilot interference, and their effect can be eliminated using a robust \textit{principal component analysis} (PCA) approach \cite{Xu2012} (adapted to estimate the dominant signal subspace from samples affected by a few strong outliers). For the geometrically consistent channel model used in \cite{gottsch2022subspace}, it is shown that after subspace projection the system performance is almost identical to the case of ``ideal but partial'' channel knowledge, i.e., to the case where each AP knows perfectly the channel vectors of all its associated users, but has only statistical knowledge of the users not associated to it.

\subsection{Aspects of Synchronization and Calibration}
\label{subsec:sync}

Different kinds of synchronization are required for distributed APs to work together. First, they need alignment in time; in principle, the requirement is no stricter than the duration of the cyclic prefix, with OFDM modulation, say. Second, they need to have a common frequency reference, and if joint reciprocity-based beamforming is to be used on DL, they also need to be jointly reciprocity-calibrated (``phase-aligned'') \cite{larsson2023phase}. 

The time synchronization is comparatively easy, while alignment in phase can be more difficult. There are three basic ways that APs can be interconnected and a phase reference can be available (or not available). A first option is to distribute the actual RF signal itself from a central unit, for example, over a (plastic) fiber, such that no local oscillators are needed at the AP. A second option is to only distribute the [baseband] data, and along with those data, distribute a low-frequency reference carrier that drives local phase-locked loop oscillators at the APs. This way, there will be no drift in the carrier frequency over time between different APs, but there will be phase noise, and there will be an unknown phase shift between the APs that needs to be measured and compensated for before starting up the system. The third option is to equip the APs with free-running oscillators; in this case, re-calibration for joint reciprocity is required every time the oscillators have drifted by more than some amount \cite{nissel2022correctly,larsson2023phase}. If the oscillators drift quickly -- a characteristic of inexpensive components -- re-calibration may be required many times every second. 

Joint reciprocity calibration can be performed \emph{over-the-air} using bidirectional measurements between APs \cite{cao2023experimental, balan2013airsync}. Such over-the-air calibration obviates the need for synchronization cables, which are considered costly, but on the other hand imposes a substantial signaling load and disrupts the TDD flow:  some panels will need to transmit calibration signals at the same time as others receive those signals. 
With multiple-antenna APs, beamforming can advantageously be performed, exploiting the array gain and avoiding unduly spreading  of interference \cite{vieralarsson_pimrc,ganesan2023beamsync}.

\subsection{What is the Role of Power Control?}

As in any wireless network, power allocation is important in CF-MMIMO to tackle the near-far effect and minimize the interference~\cite{Hien:cellfree,Nayebi2017}. Power control strategies can be implemented either in a centralized fashion at the CPUs or in a distributed fashion at each AP. Centralized implementations require a certain degree of CSI sharing over the fronthaul network but can maximize network-wide utility functions, such as sum or max-min SE. In general, these solutions are neither scalable nor low-complexity. 
Distributed implementations are more scalable as they solely rely on local channel estimates available at each AP. However, distributed solutions are usually far from  optimal.
Any non-trivial network-wide optimal power allocation is unscalable as the number of AP and/or users grows unboundedly. Distributed power allocations are heuristic, scalable and low-complexity. 

Power control strategies may be carried out at two different time scales. 
\textit{Fast} power control may be performed at the small-scale-fading time scale, i.e., within a coherence block, by fine-tuning the norm of the precoding/combining vectors. By doing so, an appropriate power allocation may control the effects of the small-scale fading. In this case, the power control coefficients (absorbed by the precoding/combining vectors) 
are functions of the channel estimates.  \textit{Slow} power control can be performed at the large-scale-fading time scale, i.e., over multiple coherence blocks, by properly setting the power control coefficients as a function of the channel statistics. Slow power control is well suited to optimization as the power control coefficients are  updated much less frequently than the channel  coherence time.   

In the UL, letting all  users transmit with full power is  quasi-optimal, in many practical cases, to maximize the sum SE when MMSE receive combining is adopted at the APs~\cite{demir2021foundations}. Therefore, attempting a power allocation based on network-wide sum rate optimization is pointless, but fairness considerations can provide an important role for power control. In the latter case, \textit{fractional} power control strategies~\cite{Nikbakht2020} are able to provide better SE to the most unfortunate users.   
In the DL, fractional power control along with MMSE transmit precoding constitutes a valid sub-optimal alternative to both sum SE maximization and max-min fairness (MMF) power control. Especially when performed in a centralized fashion, fractional power control strikes an excellent trade-off between sum SE and fairness~\cite{demir2021foundations,Nikbakht2020}. When carried out in a distributed fashion, clearly, optimizing the power allocation is beneficial regardless of the target. Sum SE maximization and  MMF power control provide significant improvements over  fractional power control, especially if MR transmission (also known as conjugate beamforming) is adopted~\cite{Interdonato_uCB}.

In the CF-MMIMO literature, MMF power control is  the policy that has received the most attention by the research community. This is mainly due to two reasons: first, because the egalitarian nature of the MMF policy fits well with CF-MMIMO. Indeed, MMF works better in CF-MMIMO than in co-located massive MIMO, wherein the cell-edge users jeopardize the performance of the cell-center users.
Second, the MMF framework can easily be extended to more general power control strategies by introducing weights that scale the performance requirements, referred to as \textit{weighted}  MMF. Weighted MMF can be also adopted to handle  \textit{user prioritization}. Indeed, SE requirements can vary from user to user. For instance, real-time application users or users with more expensive subscriptions have higher priority. In this case, a higher priority would correspond to a larger weight.  Nevertheless, MMF power control is not, in general, a recommendable solution, especially for the CF-MMIMO UL. Indeed, in some cases, MMF power control may substantially penalize UEs with good channel conditions. Moreover, introducing user-specific weights may lead to complicated UE prioritization problems. On the other hand, sum SE maximization is often the preferable utility function since it may provide excellent values of SE for the UEs with good channel conditions without breaking down the performance of UEs with poor channel conditions.

To summarize, to preserve the scalability of the system, power control implementations should be either heuristic and fully distributed, or sub-optimal and centralized but requiring only partial CSI sharing.

As an alternative to model-based optimization methods, learning-based approaches can be used to design optimal power control schemes -- potentially lowering the computational requirements,  thereby making optimal power allocation strategies appealing even for online implementation~\cite{Bjornson2020_deeplearning,ZhaoY_deeplearning}. The works~\cite{ZhaoY_deeplearning,Chakraborty2019,Nikbakht2019_deeplearning} approximately solve the MMF power allocation problem in CF-MMIMO by using techniques based on machine learning (ML). Interestingly,~\cite{Chakraborty2019} approximately solves the network-wide MMF power allocation problem by training its neural network with local CSI, whereas~\cite{Nikbakht2019_deeplearning} uses unsupervised learning approaches to solve soft max-min and max-prod power allocation problems from the channel statistics only.
The max-sum power allocation problem for the UL of CF-MMIMO systems has been addressed in~\cite{DAndrea2019_deeplearning} using artificial neural networks (ANNs) and considering only the users' locations as the input to train the neural network. Instead, a deep neural network (DNN) was designed in~\cite{Bashar2020_deeplearning}, assuming an imperfect fronthaul network.

Deep reinforcement learning (DRL) techniques are also possible  to  solve power allocation problems. However, practical uses of DNNs and/or DRL-based approaches are limited due to the large training dataset to be generated using general-purpose optimization software (i.e., needed to map input matrices of CSI to the optimal power control coefficients). 
The generation of the training dataset takes long time due to the computational complexity of the solvers, and it increases as the network gets larger. Such techniques are thus unscalable and impractical. In addition, the dataset may not be made available publicly by the network operators because of certain restrictions in terms of operator policy and security.
Importantly, sharing the dataset among multiple APs or simply transmitting such a huge amount of data to a CPU for network-wide optimization requires a fronthaul network with high capacity and makes the overhead unsustainable. Last but not least, highly dynamic propagation environments require training the neural network more frequently, which may not be affordable. To this end, unsupervised reinforcement learning approaches may be convenient as they do not require prior training and enable the system to learn ``on the fly'' relaxing the computational and the fronthauling requirements. Recently, ~\cite{Chakraborty2023} proposed a suboptimal scalable power allocation scheme for max-product SINR based on unsupervised reinforcement learning techniques, to be performed in a fully distributed fashion and in real-time, with low computational requirements.

\subsection{Massive Access}

Machine-type communication (MTC)
\cite{bockelmann2016massive,boccardi2014five} represents a paradigm
shift in wireless communication due to the diverse data traffic
characteristics and requirements on delay, reliability, energy
consumption, and security.  The vision is mMTC, where wireless connectivity is provided to a massively large number of low-complexity, low-power machine-type devices. These devices enable various emerging services in healthcare, security, manufacturing, utilities, smart homes, transportation and entertainment.

The accommodation of mMTC traffic in cellular networks brings
new fundamental challenges \cite{chen2020massive}.  Previous generations of cellular systems are designed for human-type communication which aims at high data rates using large packet sizes, typically with mobile broadband as the main intended use case. In contrast, mMTC data traffic is often UL-driven with packet sizes going down as low as a few bits. Also, typically only a tiny fraction of the devices is active at a time, because of the inherent intermittency of the traffic. This is especially the case for sensor data.  As a consequence, in the massive MTC context, the signaling overhead makes many conventional techniques almost useless and new approaches to random access and the control plane, or even the data transmission, are required. 
Furthermore, in many applications the requirements on service availability and latency are extreme, in the order of sub-millisecond end-to-end latency constraints, and reliability targets down to $10^{-8}$--$10^{-10}$.  These are unsolved challenges that call for new solution techniques.

In mMTC, only a small fraction of the devices is active at a time.  One reason for this sporadic traffic pattern is the inherent intermittency of the traffic (especially for sensor data);   the use of higher-level protocols that generate bursty traffic also contributes.  The inherent sparsity makes compressed sensing approaches a promising solution to the device detection problem \cite{choi2015downlink}. While approaches in the literature assume the availability of full or partial channel state information \cite{gao2016compressive}, in mMTC setups, the acquisition of channel state information requires the allocation of additional resources and it is also possible the assumed channel models do not capture the real characteristics of the transmission channels. Furthermore, traffic patterns are random and a priori unknown, and may be learned using appropriate learning algorithms. The combination of mMTC with massive MIMO is especially useful due to the large number of spatial degrees of freedom offered by massive MIMO.

More specifically,  if all users could be assigned orthogonal pilot sequences, conventional approaches are known to provide optimal performance. However, due to limited channel coherence, this is a physically unrealistic setup for mMTC with large numbers of devices.
A solution is to assign \emph{non-orthogonal pilots} to the users and use advanced signal processing algorithms to detect which users that are active. Activity detection with known pilots and unknown channels can be cast as an (overdetermined) linear regression problem with sparsity in the parameter vector. Thereby the regressor contains the pilots and the parameter vector contains the channel responses; the sparsity pattern corresponds to the user activities, such that the parameters are zero for inactive users. Such regression problems can be approached using compressed sensing \cite{liu2018massive}. 

A variety of algorithmic approaches have been developed. One may, for example, use approximate message passing (AMP), a low-complexity iterative algorithm proposed for general sparse regression in~\cite{donoho2009message}. For \emph{co-located} massive MIMO, this approach was studied in \cite{liu2018massive,KeMalong2020}. For CF-MMIMO, it has been used in  \cite{Han2014,guo2021joint,jianan2022,Chen_2021}. The approach taken in \cite{jianan2022,Chen_2021} is to take independent soft decisions at different APs, and fuse
these decisions by adding up log-likelihood ratios.

Another approach to activity detection is to assume a distribution of the channel vectors (Gaussian, typically) and parameterize the problem in terms of the variances of these distributions. Given this parameterization, one can formulate a maximum-likelihood problem and estimate the variances. Users for which the variance estimate is nonzero are declared to be active. The estimation problem can be solved using, for example, coordinate descent \cite{fengler2021non}. This method tends to outperform \textit{approximate message passing} (AMP) for activity detection in co-located MIMO. 
For CF-MMIMO, the approach cannot be directly used, since the fading statistics from a given user is different to the different APs. Extensions of the method,  that rely on AP clustering together with appropriate approximations,
have been developed \cite{ganesan2021clustering}.

\subsection{How to Implement Dynamic Scheduling?}

Early works on CF-MMIMO assumed single-antenna APs 
($N = 1$) and $M  >  K$ \cite{Hien:cellfree}.
Recognizing that the placement of ``more APs than users'' is hardly justifiable from an operator deployment cost viewpoint, more recent works  have considered a more realistic AP/user density 
regime $M < K < NM$ with $N > 1$  (e.g., see \cite{Bjornson_Sanguinetti_Scalable_TCOM2020,demir2021foundations,miretti2022team, gottsch2022subspace,chen2022}). 
In these works, the $K$ users are all simultaneously active and the system performance is studied in 
terms of the per-user {\em ergodic rates} (e.g., see \cite{ngo16,Bjornson_Sanguinetti_Scalable_TCOM2020,demir2021foundations,miretti2022team,gottsch2022subspace}).  However, the achievability of ergodic rates assumes that coding can be performed over a sufficiently large sequence of independent channel fading states, implying continuous transmission over many time-frequency slots. 
This assumption may be incompatible with dynamic scheduling and per-slot coding/decoding, 
as well as with low-latency requirements, which are  key features of 5G \cite{itu2017requirements, Durisi2016}.
In addition, the number of actual users in the system $K_{\rm tot}$ is in general much larger than 
the total number of system antennas $NM$. For example,  consider a system where the total bandwidth is partitioned into 
$100$ frequency resource blocks (RBs) per time slot, serving a total number of $10,000$ users with $M = 20$ APs with $N = 10$ antennas each (e.g., see the real-world deployment in \cite{Forenza2015}). 
Every active user is allocated a block of $F = 10$ RBs in frequency to achieve a certain level of frequency diversity. 
Thus, the scheduler dynamically chooses for every slot a set of $K \approx \frac{NM}{2} = 100$
users out of $K_{\rm tot} = 1,000$ per RB in order to 
exploit the   spatial degrees of freedom offered by the system.\footnote{It is important to clarify that $K$ represents the number of active users actually scheduled within a given time/frequency resource. As a rule of thumb, it has been noticed that, having 
a number of active users ($K$) roughly equal to half of the total number of system antennas yields
a good tradeoff between the total SE and the instantaneous rate of each active user \cite{gottsch2023fairness}.}  
A good scheduler should choose a number of active users $K$ out of $K_{\rm tot}$ in each slot such that
the overall system SE is high, the instantaneous rate per scheduled user on a single slot is not too low, 
and the long-term average throughput rate of all $K_{\rm tot}$ users is ``fair'', i.e., 
each user in the system gets its fair share of the total  (time-averaged) SE. 
The scheduler must allocate an ``instantaneous'' rate to each active user since encoding/decoding is performed 
block by block, i.e., coding over a virtually infinite sequence of fading states is not possible.  
In this case, the instantaneous rate must be scheduled according to the notion of {\em information outage rate} (e.g., see \cite{biglieri1998fading}), where a non-vanishing block error probability is taken explicitly into account.

The problem of dynamic scheduling in user-centric CF-MMIMO networks where the number of total users
is much larger than the number of simultaneously active users on each time-frequency RB has not been widely investigated 
and remains quite open. The recent work \cite{gottsch2023fairness} formulates the problem as 
network utility maximization (NUM) where the utility function is a concave $\alpha$-fairness function \cite{mo2000fair}
of the users' throughput rates (i.e., long-term averaged rates), and the scheduler adapts its priority weights according to
the Lyapunov drift-plus-penalty (DPP) approach (e.g., see \cite{neely2008fairness,shirani2010mimo}). 
The DPP approach solves at each scheduling decision interval a constrained weighted sum-rate maximization problem based on the individual users' outage rates, where the constraint is expressed as conflict graph that prevent the simultaneous scheduling of 
users with strong mutual pilot contamination. Systems with tens of thousands of users and hundreds of antennas are demonstrated, verifying the experimental results posted  by \cite{Forenza2015}.

\subsection{Emerging Cell-Free Wireless Access Infrastructures: The State-of-the-Art in the Industry} 

CF-MMIMO requires a widespread and costly architecture to implement the user-centric philosophy, accurate synchronization and coordination among the APs to carry out the joint coherent transmission/reception, and practical resource allocation schemes to capitalize on the macro-diversity gain, while preserving the network scalability. 

Luckily, there are emerging low-cost and flexible solutions for cell-free network deployments proposed by industry, such as \textit{pCell}~\cite{Forenza2015, Perlman2015} by Artemis Networks, and  \textit{Radio Stripes}~\cite{interdonato2019ubiquitous,patentRadioStripeUS} by Ericsson  that promise to achieve to the promises of CF-MMIMO:  outstanding SE, reliable communication, network scalability,  link robustness, low latency and low power consumption both at the user and at the AP. Another recent physical-layer wireless access concept,  \textit{RadioWeaves}~\cite{VanderPerre2019},  combines the large-scale intelligent surfaces and CF-MMIMO concepts, mainly targeting indoor deployments.

\begin{figure}[!t]
\centering
\resizebox{\columnwidth}{!}{\input{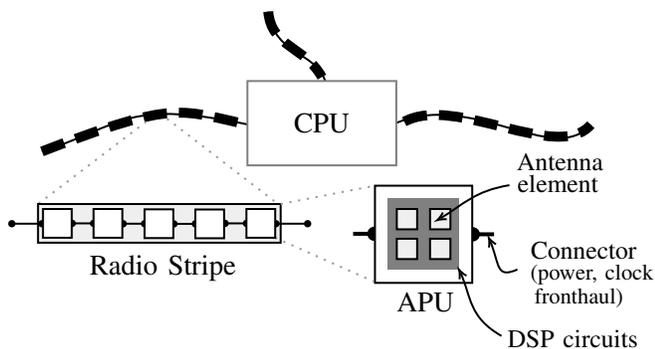}}
\caption{CF-MMIMO deployment with the Radio Stripes. The APUs are serially integrated into stripe and consist of circuit-mounted chips necessary for the baseband processing and antenna elements. A shared connector provides power supply, synchronization and data transfer.}
\label{fig:stripes}
\end{figure}

\medskip
\textbf{Radio Stripes and Sequential Signal Processing: }
The basic principle of the Radio Stripes is to fuse into a single entity the antenna elements, the baseband processing circuitry and the fronthaul network.
Specifically, the antenna processing units (APUs) are serially located inside the protective casing of a stripe, which also provides power supply, synchronization and data transfer through a shared connector, as in the example illustrated in Figure~\ref{fig:stripes}. The APU consists of the antenna elements and the circuits for digital signal processing (DSP), namely circuit-mounted chips accommodating phase shifters, modulators, filters, power amplifiers, as well as analog-to-digital and digital-to-analog converters.

A Radio Stripe system hinges on a compute-and-forward architecture wherein the signal processing occurs sequentially. At the ends of a stripe a CPU collects the cumulative processed signal and performs network tasks in a centralized fashion.
The signal processing is serialized, that is the receive/transmit signal processing of an APU is carried out right next to itself. When transmitting, each APU receives a superposition of multiple input data streams from the previous APU via the shared connector, applies its precoding vectors and finally transmits the resulting signal through its antenna elements. When receiving, each APU first processes the radio signal received by its antennas, then  combines the resulting streams with the data streams received from the previous APU, and finally   sends the resulting signal through the fronthaul to the next APU. 
This combination of signals might simply be a per-stream addition operation. 

Such sequential signal processing over a serial fronthaul network has been recently demonstrated to provide an optimal SE in the UL~\cite{Shaik2021} and a nearly-optimal SE in the DL~\cite{Miretti2021,Miretti2021_team,miretti2022team}. Hence, CF-MMIMO can be efficiently deployed by using a sequential topology without loss in performance but with much lower fronthaul requirements compared to centralized (i.e., star-like) implementations.
Indeed, the work~\cite{Shaik2021} finds the optimal receive combining matrices within the class of sequential receivers that jointly maximize the SE and minimize the mean square error. This is attained by using a linear MMSE receiver. Importantly,~\cite{Shaik2021} shows analytically that, with UL sequential processing, the data estimate computed at the last APU is equivalent to that obtained by centralized processing. However, the fronthaul capacity per stripe grows with the number of users in the former case and with the number of APs in the latter case. Since in a CF-MMIMO system we have more APs than active users (by definition) then the percentage of fronthaul resources saved by the sequential processing becomes considerable. Similarly, in the DL the TMMSE precoding for sequential processing, proposed in~\cite{Miretti2021,Miretti2021_team,miretti2022team}, provides nearly-optimal SE and minimizes the mean-square error under partial CSI sharing. In particular, unidirectional TMMSE, which is obtained by sharing the CSI in only one direction over the serial fronthaul, is a promising intermediate solution for supporting network-wide interference management when centralized precoding is expensive.

Besides the advantages described above, the Radio Stripe system facilitates a practical, flexible and cheap CF-MMIMO deployment by using sequential processing over a serial network topology. The benefits include diverse practical aspects: $(i)$ ease of deployment and cable routing; $(ii)$ cabling is cheaper than any other topology, e.g., star-like, tree, mesh. Only two links between any pair of APs are needed for sending/receiving the signals; $(iii)$ node failures can be tolerated by routing mechanisms thereby increasing robustness and resilience; $(iv)$ components are low-power and distributed, thus heat-dissipation is less alarming than ``packed'' solutions; $(v)$ while conventional APs are bulky, radio stripes enable non-invasive deployments.
On the other hand, the Radio Stripe system cannot be the ideal solution for any large-scale distributed deployment. For instance, the embodiment depicted in Figure~\ref{fig:stripes} subsumes a combined digital-electrical fronthaul interface technology such as \textit{IEEE 802.3bt Power-over-Ethernet (PoE)}~\cite{PoE2018} which is able to support a data transfer rate up to 10 Gbps. Such a limited capacity makes the Radio Stripe system built on PoE suitable for sub-6 GHz frequencies~\cite{Gustavsson2021}. Conversely, optical fronthaul interface technologies (e.g., analog radio-over-fiber~\cite{Apostolopoulos2018} and $\Sigma\Delta$-over-fiber~\cite{Sezgin2018}) are required to support the larger data rates achievable by CF-MMIMO systems operating at the high frequency bands (e.g., millimeter Wave and THz).

\textbf{pCell by Artemis Networks\footnote{Disclosure: J. G. Andrews and G. Caire are on the Technical Advisory Board of Artemis.}:} The pCell technology \cite{Forenza2015} uses multiple small-cell BSs to synthesize a ``personal'' cell (i.e. a ``pCell'') for each user in the coverage area.  This is the essence of of ``cell-free'' network MIMO: each user has effectively its own custom coverage area.   Artemis's pCell technology follows the same principles as described for CF-MMIMO: TDD operation, pilot-based UL training, and centralized digital precoding for the DL and digital combining for the UL. 
This is attained via a fully connected (via fiber and high rate Ethernet) and centralized CRAN architecture compatible with the LTE and/or 5G NR standards.   A key enabler for pCell's performance is the attainment of high precision CSI and synchronization across the network, which is essential for attaining constructive superposition of the signals from the nearby APs at each user.  It is fundamentally difficult in pCell, or any such CF-MMIMO system, to support substantial mobility, given the CSI and user-centered synchronization requirements.

\textbf{RadioWeaves: }
The RadioWeaves wireless access infrastructure was conceived in~\cite{VanderPerre2019}, and is primarily intended for indoor environments. The idea is to integrate distributed antennas into naturally occurring and man-made structures and objects, for example the ceiling or walls of buildings, or even furniture. Such integration could be made aesthetically appealing, or even invisible to the bare eye.  Given the large apertures that can be obtained this way, extraordinary levels of multiplexing, spatial diversity, energy efficiency, link reliability and connectivity can be achieved \cite{ganesan2020radioweaves}.
The general communication theory developed for CF-MMIMO applies to RadioWeaves in the same way as it does for Radio Stripes and pCell. 

Importantly, RadioWeaves, and CF-MMIMO more generally, naturally operates in the [geometric] near-field of the ``super-array'' collectively constituted by all antenna panels together. In fact, the topology of the actual antenna deployment hardly matters at all. With reciprocity-based beamforming, the physical shape of the actual beams, and grating lobe phenomena in particular, becomes irrelevant. If anything, given a set of antennas, it is advantageous to spread them out over as a large aperture as possible. One should avoid placing antennas closer than half a wavelength together: such dense packing of antennas is almost never meaningful, as sampling points $\lambda/2$-spaced apart captures essentially all the degrees of freedom of the field. Putting the antennas closer results in coupling effects that are usually of more harm than benefit.

One important technical challenge in constructing RadioWeaves system is, down-to-earth, to handle the vast amounts of baseband data, and process them in real time. Another is time and phase synchronization of distributed MIMO arrays (see Section~\ref{subsec:sync}). A third is initial access, covering space uniformly with system information signals, and waking up sleeping devices. A fourth is energy-efficiency, at all levels in the network. A fifth is the integration of service of energy-neutral devices that communicate via backscattering. CF-MMIMO naturally offers the infrastructure for that, permitting simultaneous transmission and reception from different panels in a bistatic setup; however, these activities break the TDD flow and must be carefully integrated into the workings of the system.


\subsection{Other Design Aspects of  Ultra-Dense CF-MMIMO} 
\label{subse:recipe}

\medskip
\subsubsection{Low-complexity processing}\label{subsec:low-complexity}
When designing interference-suppression-based precoding and combining schemes, the complexity of implementing the beamforming vectors mainly lies in the computation of the pseudo-inverse matrix, hence on its size. In CF-MMIMO, due to the distributed network topology, the dynamic range of the interference from/to different APs may be  large. Hence, there is no need to design a precoder (combiner) that suppresses all the interference contributions, but rather only the most significant ones, in terms of both inter-cell and intra-cell interference. This is the basic principle behind, for instance, the \textit{partial zero-forcing} (PZF) scheme proposed in~\cite{Interdonato_PZF_2020,ZhangJ2021} and the \textit{local partial} MMSE (LP-MMSE) in~\cite{demir2021foundations}. These are fully distributed precoding schemes that provide
an adaptable trade-off between interference cancellation and boosting of the desired signal, with no additional fronthaul overhead. These schemes only suppress the most significant intra-AP interference contributions; hence the inter-AP interference is not tackled since there is no cooperation among APs.
A semi-distributed variant of the PZF that also partially suppresses the inter-AP interference by exchanging a limited amount of CSI over the fronthaul network was proposed in~\cite{Du2021}. With this scheme, only a subset of APs uses network-wide ZF, while the rest uses MR transmission, thus the dimension of the pseudo-inverse and the amount of the exchanged data is determined by the cardinality of the ZF set. Such a joint MR transmission and ZF (JMRZF) precoding scheme converges to the performance of network-wide ZF as the number of APs included in the ZF set grows, at the expense of an increased computational complexity and fronthaul overhead.

\medskip
\subsubsection{Low-dimensional CSI exchange} \label{subsec:low-dimensional-CSI}
Although the DCC framework allows to confine the data sharing among a subset of APs, hence relieving the fronthaul network, CSI sharing between the APs needs to be performed within much tighter time constraints, and hence may dominate the fronthaul overhead. A viable approach to limit the CSI sharing consists in implementing cooperative transmission strategies only among a few APs. Alternatively, each AP may operate on the basis of possibly different estimates of the global channel state obtained through some arbitrary CSI acquisition and sharing mechanism. For instance, in~\cite{miretti2022team} a distributed precoding design is proposed, coined \textit{Team} MMSE precoding, that generalizes the network-wide MMSE precoding to distributed operations based on transmitter-specific CSI.
On the other hand,~\cite{Atzeni2021} proposes a distributed framework for cooperative precoding in CF-MMIMO systems based on a novel \textit{over-the-air} (OTA) signaling mechanism that entirely eliminates the need for fronthaul signaling for CSI exchange.
Importantly, the amount of OTA signaling does not scale with the number of APs or users and there are no delays in the CSI exchange among the APs. These practical benefits come at the cost of extra UL signaling overhead per bi-directional training iteration, which, however, results in a minor performance loss with respect to the distributed precoding design via fronthaul signaling.

\medskip
\subsubsection{Robustness against hardware impairments}\label{subsec:hardware-impairments}

To keep down the costs of deploying an UDN, it is desirable to 
use inexpensive hardware.
Nevertheless, low-quality hardware may introduce distortion that causes significant system performance losses. 
The distortion  that arises due to the use of non-ideal hardware  includes phase noise, carrier frequency offsets resulting in inter-carrier interference, and non-linearities in the transceivers. 
Nonlinearities represents one of the most important and serious effects.
Two sources of nonlinearities are the quantization in the ADCs (in UL), and the power amplifiers (on DL). 
Power amplifier nonlinearities, specifically, can cause significant out-of-band emissions, which in turn can affect systems operating on neighboring frequency bands. Typically, communication standards specify strict limits on out-of-band emissions.

Concentrating on nonlinearities, a universally valid conclusion for both cell-free and co-located MIMO is that the distortion stemming from nonlinearities   is a deterministic function of the signals. One consequence   is that in most cases of practical relevance, the distortion is correlated among the antennas. For example, on DL, under line-of-sight propagation conditions this distortion is radiated in specific directions: in the single-user case the distortion is radiated into the same direction as the signal of interest, and in the two-user case the distortion is radiated into two other directions \cite{larsson2018out}. In fact, even in non-line-of-sight,  distortion also coherently combines at the user location(s), although in that case ``direction'' no longer has a meaning.
This specifically implies that  out-of-band emissions are amplified by the array gain in the same way as the desired signals. These phenomena were analyzed in detail in \cite{mollen2018spatial,larsson2018out} for co-located MIMO and in \cite{Rottenberg2021,rottenberg2022z3ro} for CF-MMIMO. Not much is different, qualitatively in terms of conclusions, between cellular and CF-MMIMO; for details, see  the cited papers.
Similarly, on UL, quantization noise in the ADCs becomes correlated among the antennas, causing sensitivity to blocking (high-power signals in neighboring bands).
 
In general, hardware impairments on UL can be mitigated to some extent using appropriate receiver processing algorithms; on DL this is more difficult.
While DL precoders can be adapted to combat the effects of nonlinearities to some extent \cite{rottenberg2022z3ro,feys2023self}, observing strict out-of-band emission requirements can be challenging. 

Under some circumstances, if the fading pattern changes very quickly with time (e.g. Rayleigh fading) and beamforming is performed simultaneously to many users, a certain time-averaging  can be justified and the hardware distortion can be approximated as uncorrelated across the antennas. This assumption, which is valid only in special circumstances,  has been popular in some literature 
\cite{ZhangJ2018_HWimpairments,Elhoushy2020_HWimpairments,Masoumi20200_HWimpairments,ZhengJ2020_HWimpairments}. However, it is important to realize that in many cases, for example, a static line-of-sight channel, this approximation has no physical basis.  If out-of-band radiation (the most difficult and important hardware impairment on DL) is of concern, then to meet out-of-band emission requirements one has to account for the worst case, in which case the uncorrelatedness-approximation is inapplicable and the rigorous analyses of \cite{mollen2018spatial,larsson2018out,Rottenberg2021} must be used.

\section{Conclusions} \label{sec:conclusions}

In an age where wireless communications is called upon to provide high-speed, reliable connectivity anywhere, anytime, to anyone, emerging heterogeneous users' requirements pose novel design and implementation challenges. This has  led academic and industrial research to look beyond the cellular paradigm that characterizes the current mobile systems. In this regard, CF-MMIMO is deemed to be a key physical layer technology for beyond-5G systems. It combines the efficiency, practicality and scalability of massive MIMO with a ultra-dense distributed deployment, and coordinated multipoint processing, with a user-centric approach. The result of this combination is a uniform great quality of service to every user.
However, CF-MMIMO requires a widespread and costly architecture, accurate synchronization and coordination among the APs, and practical resource allocation.

We provided a comprehensive survey on ultra-dense CF-MMIMO touching different aspects: practical network infrastructure, signal processing, capacity bounding techniques, massive access, resource allocation, synchronization and calibration. Importantly, we discussed key open questions related to practical implementations of CF-MMIMO, particularly under ultra-dense scenarios and realistic constraints. Throughout  the article, we pointed out the key ingredients that a CF-MMIMO system should have, thereby providing clear directions for future research and development of this promising technology. In addition to the key questions addressed in this paper, there are still many interesting research questions requiring future comprehensive studies, such as 1) At what level of densification is CF-MMIMO truly
beneficial? When is it not really needed?; 2) How to handle high mobility in UDNs with CF-MMIMO?; 3) CF-MMIMO for communication with backscattering devices;  4) How CF-MMIMO works (including channel estimation, scheduling procedures, signal processing at the users) with multiple-antenna users;
 and 5) How to combine CF-MMIMO with other advanced technologies such as reconfigurable intelligent surface, near-field transmissions, and AI technologies.


\bibliographystyle{IEEEtran}
\bibliography{refs}

\end{document}